\documentclass[oldversion]{aa}
\usepackage{color}
\usepackage{amsmath,amsbsy}

\newcommand{\se}[1]{\mbox{Sect.\ \ref{sec:#1}}}

\newcommand{\Se}[1]{\mbox{Section\ \ref{sec:#1}}}
\newcommand{\eq}[1]{\mbox{Eq. (\ref{eq:#1})}}
\newcommand{\Eq}[1]{\mbox{Equation\ (\ref{eq:#1})}}
\newcommand{\fg}[1]{\mbox{Fig.\ \ref{fig:#1}}}
\newcommand{\eqs}[2]{Eqs.\ \ref{eq:#1} and \ref{eq:#2}}

\newcommand{\fgs}[2]{Figs.\ \ref{fig:#1} and \ref{fig:#2}}

\newcommand{\Fg}[1]{\mbox{Figure\ \ref{fig:#1}}}
\newcommand{\Tb}[1]{\mbox{Table\ \ref{tab:#1}}}
\newcommand{\app}[1]{\mbox{Appendix\ \ref{app:#1}}}
\newcommand{\ie}{i.e.,}

\newcommand{\eg}{e.g.,}
\newcommand{\etc}{etc.}
\newcommand{\cf}{cf.}
\newcommand{\sumi}{(i)}
\newcommand{\sumii}{(ii)}

\usepackage{natbib,graphicx}
\usepackage[varg]{txfonts}
\title{The effect of gas drag on the growth of protoplanets}
\subtitle{Analytical expressions for the accretion of small bodies in laminar disks}
\author{C.W. Ormel \and H.H. Klahr}
\institute{Max-Planck-Institut f\"ur Astronomie, K\"onigstuhl 17, 69117, Heidelberg, Germany;
          \email{[ormel,klahr]@mpia-hd.mpg.de} }
\abstract{ Planetary bodies form by accretion of smaller bodies.  It has been suggested that a very efficient way to grow protoplanets is by accreting particles of size $\ll$km (\eg\ chondrules, boulders, or fragments of larger bodies) as they can be kept dynamically cold.  We investigate the effects of gas drag on the impact radii and the accretion rates of these particles.  As simplifying assumptions we restrict our analysis to 2D settings, a gas drag law linear in velocity, and a laminar disk characterized by a smooth (global) pressure gradient that causes particles to drift in radially.  These approximations, however, enable us to cover an arbitrary large parameter space.  The framework of the circularly restricted three body problem is used to numerically integrate particle trajectories and to derive their impact parameters.  Three accretion modes can be distinguished: \textit{hyperbolic encounters}, where the 2-body gravitational focusing enhances the impact parameter; \textit{three-body encounters}, where gas drag enhances the capture probability; and \textit{settling encounters}, where particles settle towards the protoplanet.  An analysis of the observed behavior is presented; and we provide a recipe to analytically calculate the impact radius, which confirms the numerical findings.  We apply our results to the sweepup of fragments by a protoplanet at a distance of 5 AU.  Accretion of debris on small protoplanets ($\lesssim$50 km) is found to be slow, because the fragments are distributed over a rather thick layer.  However, the newly found settling mechanism, which is characterized by much larger impact radii, becomes relevant for protoplanets of $\sim$10$^3$ km in size and provides a much faster channel for growth.  }
\newcommand{\figw}{0.48\textwidth}


\newcommand{\arrayiii}[3]{\left( \begin{array}{c} #1 \\ #2 \\ #3 \\ \end{array} \right)}
\newcommand{\arrayii}[2]{\left( \begin{array}{c} #1 \\ #2 \\ \end{array} \right)}

\begin{document}
\maketitle
\section{Introduction}
\label{sec:intro}
We consider how gas drag affects the collision rates between a big body -- a planetesimal or protoplanet -- and small particles, \eg\ dust, chondrules, or boulders.  Although the core accretion model \citep{PollackEtal1996,HubickyjEtal2005} in its initial stages, \ie\ before the formation of a $\sim$10 Earth mass ($M_\oplus$) core, concerns the accumulation of solid bodies, the role of the gas cannot be overstated.  In the early phases of planet formation -- the growth of dust to planetesimals -- the gas damps the velocities of small particles.  Initially, the (relative) velocities between particles are tiny and this is the reason why dust grains can coagulate due to intermolecular forces \citep{DominikTielens1997,BlumWurm2000} -- an effect much harder to envision in the diffuse interstellar medium or even in molecular clouds.  In this stage mechanisms that induce a relative velocity among the dust particles include Brownian motion, settling, radial drift, and turbulent motions.  The latter three are all functions of the particle's stopping time, a measure of how well particles couple to the gas.  With increasing size (or, more correctly, increasing mass-to-surface area) particles couple less well to the gas and relative velocities increase, culminating in the so called meter-size barrier, which, at our current level of understanding, can best be overcome by the combined efforts of turbulent concentration and gravitational collapse \citep{JohansenEtal2007,JohansenEtal2009,CuzziEtal2010}.

Gas drag also affects the collisional behavior at a much later stage, when protoplanets accrete planetesimals of perhaps $\sim$$1-10^2$ km in size.  The collisional cross section between these big bodies is increased by gravitational focusing, \ie\ the body can accrete particles at a cross section larger than its geometrical cross section due to gravitational deflection \citep{Safronov1969,WetherillStewart1989,GreenzweigLissauer1990,GreenzweigLissauer1992}.   This effect, however, is very sensitive to the velocity $v_a$ at which the bodies approach: if $v_a$ is too large, the focusing vanishes.  In planetesimal accretion theory it is expected that a protoplanet will excite the random motions (eccentricities and inclinations) of the bodies it is accreting from, leading to a self-regulated accretion behavior, which slows down the growth \citep{IdaMakino1993,KokuboIda1998,OrmelEtal2010}.  Gas drag can provide some relief since, by damping the random motions of the planetesimals, the gravitational focusing is kept large.   Moreover, the capture probability of planetesimals is also significantly increased when (proto)planets are surrounded by atmospheres \citep{InabaIkoma2003,TanigawaOhtsuki2010} -- again, gas drag is the mechanism that facilitates their accretion.  Still, it is unclear if these effects are sufficient to overcome the timescale problem, \ie\ to grow protoplanets to $\sim$10 $M_\oplus$ within the time the gas disk dissipates ($\sim$10$^6$ yr); see \citet{LevisonEtal2010} for a recent review.

Due to the dynamical heating of planetesimals, planetesimal-planetesimal collisions may become disruptive, producing smaller planetesimals or even fragments \citep{WetherillStewart1993,LeinhardtEtal2009}.  These fragments can be kept dynamically cold, \eg\ by mutual collisions or by gas drag.  The accretion then takes place at low $v_a$ -- the shear-dominated regime -- which is very favorable for growth \citep{GoldreichEtal2004}.  The generation of large amounts of fragments therefore can significantly boost accretion .  In particular, the accretion rate in the two dimensional (interactions are confined to a plane), \textit{gas-free}, three-body regime (including the gravity of the central star) is derived by a number of studies to be
\begin{equation}
  \left(\frac{dM}{dt}\right)_\mathrm{gf} \approx 11 \alpha_p^{1/2} R_h v_h \Sigma
  \label{eq:dMdt}
\end{equation}
(\eg\ \citealt{IdaNakazawa1989,GreenbergEtal1991,Rafikov2004,Weidenschilling2005,OrmelEtal2010i}; the numerical constant is adopted from \citealt{InabaEtal2001}), where $\Sigma$ is the density in solids, $R_h$ the Hill radius,
\begin{equation}
  R_h = a \left( \frac{M_p}{3M_\star} \right)^{1/3},
  \label{eq:Rh}
\end{equation}
$v_h$ the Hill velocity, $v_h = R_h \Omega$, $a$ the semi-major axis, $\Omega$ the corresponding orbital frequency, $M_p/M_\star$ the ratio between the mass of the protoplanet and the central star, and $\alpha_p$ the ratio between the protoplanet radius and the Hill radius, $\alpha_p = R_p/R_h$. \Eq{dMdt} is often used in statistical models for the accretion rate \citep[\eg][]{InabaEtal2001,Chambers2006,BruniniBenvenuto2008,Chambers2008,KobayashiEtal2010}.  It represents a fast accretion rate. \citet{KenyonBromley2009}, applying such a fragmentation-driven accretion scenario, calculate that the core formation process can be completed within $10^6$ yr.%, a conclusion that reflects \citet{GoldreichEtal2004} shear-dominated growth.

How would gas drag affect these conclusions; \ie\ does the rather large accretion rate of \eq{dMdt} also materialize in the presence of gas drag?  Qualitatively, two directions can be envisioned.  On the one hand, the dissipative nature of the drag will enhance the collision (impact) radius, like in the case of a dense atmosphere.  Conversely, strong particle-gas \textit{coupling} will suppress the accretion efficiency since the gas after all is not accreted but flows past the object (until the point where it has become more massive than 10 $M_\oplus$ and gas runaway accretion kicks in).  It is \textit{a priori} unclear which aspect of the drag -- the coupling or the dissipation -- will turn out to be the more important. %Also, there is the effect of radial drift.  \citet{LevisonEtal2010} generally found that fragments drifted in before the protoplanet could accrete them.

To address these questions we include gas drag as an additional force to the restricted 3-body problem that has been previously used in calculating accretion rates in gas-free systems (or in systems where gas can be neglected; \citealt{PetitHenon1986,IdaNakazawa1989}).  Using appropriate scaling behavior, we show, in \se{sketch}, that the system of equations containing all the physics can be restated into two dimensionless parameters: the dimensionless headwind velocity $\zeta_w$ that the protoplanet experiences and the dimensionless stopping time (Stokes number, $\mathrm{St}$) of the particle.  Our setup is idealized in the sense that we assume a steady gas flow of constant density (\ie\ no pressure fluctuations or atmospheres), a drag law linear in velocity (applicable to small particles), and only consider drift motions of particles.

After having outlined our setup in \se{sketch}, \se{geo} considers the geometrical limit, in which the 2 body interaction is absent or can be ignored.  In \se{orbitints} we perform an extensive parameter study to obtain the impact parameters as function of the relevant dimensionless quantities.  \Se{model} presents an analytic model to obtain the impact radii and accretion rates from first principles, which we compare to our measured values. \Se{signif} illustrates the significance of our result by calculating the protoplanet growth timescale in which we apply a correction to account for the scaleheight of the particle layer.  We discuss limitations of our results and summarize in \se{discuss}.

\section{Sketch of problem and approach}
\label{sec:sketch}
\subsection{Definition of impact radius}
\label{sec:gas-free}
In this study we will calculate both numerically and analytically impact radii, $b_\sigma$.  In 3D systems, the collision rate $dM/dt$ is the product of the velocity at which the bodies approach each other, the approach velocity, $v_a$, the cross section for collisions, $\sigma$, and the volume density in solids $\rho$ that are accreted, $dM/dt = \rho \sigma v_a$. In 2D configurations the vertical dimension is lacking and we define 
\begin{equation}
  \left(\frac{dM}{dt}\right)_\mathrm{2D} \equiv 2b_\sigma v_a \Sigma = P_\mathrm{col} \Sigma,
  \label{eq:dMdt-2}
\end{equation}
%$dM/dt = \Sigma (2b_\sigma) v_a$.   The collision rate of \eq{dMdt} is then written as
%\com{Say that $N_sP_\mathrm{col}$ is the number of particles that collide per second}
where $P_\mathrm{col}\equiv2b_\sigma v_a$ is the specific collision rate \citep[\cf][]{NakazawaEtal1989}.  In the drag-free regime we indicate the impact radius $b_\sigma$ by $b_\mathrm{gf}$.  Although we primarily focus on 2D-configurations, \se{signif} considers a 3D extension in which we apply the derived $b_\sigma$ also for the vertical dimension.

In the gas-free regime particles enter the Hill sphere from orbits both interior and exterior to that of the planet, see \fg{sketch}.  Therefore, $b_\mathrm{gf}$ is associated with the lengthscale over which particles impact for one of these branches.  However, particles can only enter at specific intervals, $1.7R_h<|b|<2.5R_h$ \citep[\eg][]{GreenbergEtal1991};  particles on impact parameters $|b|<1.7R_h$ move on horseshoe orbits that do not enter the Hill sphere.  The approach velocity $v_a$ for the 3-body regime is defined as the average shear velocity ($3b\Omega/2$) over the above interval, \ie\ 
\begin{equation}
  v_a \equiv \frac{1}{2.5R_h - 1.7R_h}\int_{1.7R_h}^{2.5R_h} \frac{3\Omega b}{2}\ \mathrm{d}b = 3.2 v_h.
  \label{eq:va-gf-def}
\end{equation}
Using $v_a = 3.2v_h$ and equating \eq{dMdt} with \eq{dMdt-2} gives
\begin{equation}
  b_\mathrm{gf} = 1.7 \alpha_p^{1/2} R_h
  \label{eq:bgf}
\end{equation}
as the (effective) impact radius for accretion in the 2D gas-free regime.  Note that since $\alpha_p \ll 1$, $b_\mathrm{gf} \ll R_h$, which signifies that not every particle that enters the Hill sphere will collide.  For this reason we distinguish between $b_\mathrm{app}=2.5$, the impact parameter at which particles approach (which is related to $v_a$), and $b_\sigma$, the impact radius that enters in the expression for the collision rate $P_\mathrm{col}$.  The fact that $b_\sigma\neq b_\mathrm{app}$ is peculiar to the three-body regime, where the gravity of the central star becomes important.
\begin{figure}
  \centering
  \includegraphics[width=0.4\textwidth]{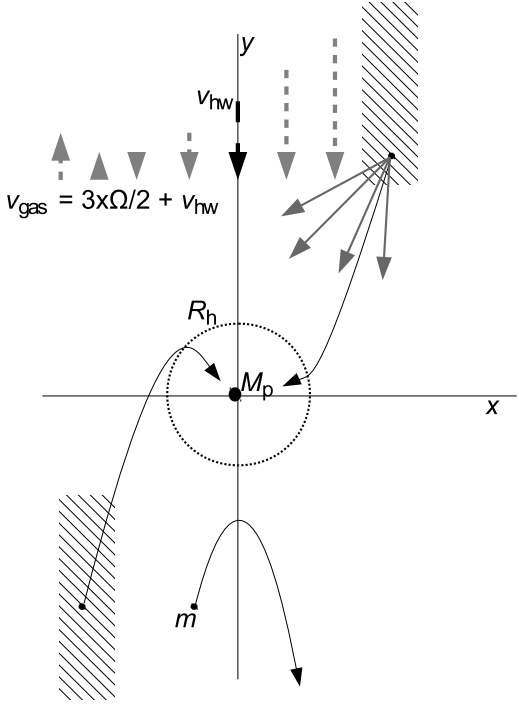}
  \caption{\label{fig:sketch}Sketch of particle trajectories in the comoving frame.  We consider the motion of the third (test) particle $m$ in the comoving frame of the second body ($M_p$, the planet) while including the gravity of the central star.  In the gas-free limit zero-eccentricity particles can enter the Hill sphere from both the first and the third quadrant (black curves) but only from specific impact parameters indicated by the hatched regions.  Particles arriving at closer impact parameters move on horseshoe orbits. The magnitude and direction of the gas velocity $\mathbf{v}_\mathrm{gas}$ as seen from the comoving frame is indicated by the dashed arrows.  Particle trajectories including gas drag (solid gray arrows) can be anything depending on the properties of the particle and the gas.}
\end{figure}

%treats the interaction between two bodies and includes the gravitational perturbation of the central object (the sun), assuming $M_1=M_\star \gg M_2, M_3$.  In the circularly restricted three-body problem we furthermore assume that the second body (the protoplanet or asteroid) moves on a circular orbit.\com{check?}  We then consider the motion of the third body as it approaches the protoplanet from an arbitrary direction.
\subsection{The circularly restricted three body problem modified by gas drag}
We briefly review the circularly restricted three body problem using the framework of Hill's equations \citep{Hill1878} and include a drag term.  The restricted three body problem assumes that the mass of the third body ($M_3$) can be neglected with respect to the masses of the other two bodies, $M_1, M_2$.  Furthermore, it is assumed that the orbits of the bodies are confined to a single plane and that these are circular for the two massive bodies.  In our case the first body is the central star ($M_\star$), the second the (proto)planet ($M_p$), and the third the (test) particle $m$.  We then consider the motion of the test particle in a coordinate system centered on and rotating with the motion of the planet.  The resulting equations of motions for $m$ in such a frame rotating with angular frequency $\Omega_0$ read
\begin{equation}
  \frac{d\mathbf{v}}{dt} = \mathbf{F} -2\Omega_0 \times \mathbf{v} + \Omega_0^2 \mathbf{r},
  \label{eq:3body-1}
\end{equation}
where $\mathbf{r} = (x,y,z)$ are the coordinates in the comoving frame, $2\Omega_0 \times \mathbf{v}$ is the Coriolis acceleration and $\Omega_0^2 \mathbf{r}$ the centrifugal acceleration.  The force per unit mass, $\mathbf{F}$, acting on the third body consist of the solar gravity, $\mathbf{F}_\mathrm{sun} = \Omega^2 \mathbf{a}$, the 2-body force with the protoplanet, $\mathbf{F}_\mathrm{2b} = GM\mathbf{r}/r^3$, and the drag force with the gas $\mathbf{F}_\mathrm{drag}$.  Expanding the $\mathbf{F}_\mathrm{sun}$ term around $a_0$ enables us to linearize \eq{3body-1} to obtain
\begin{equation}
  \frac{d\mathbf{v}}{dt} = \arrayiii{2\Omega_0 v_y + 3\Omega_0^2 x}{-2\Omega_0 v_x}{-\Omega_0^2 z} - \frac{GM}{r^3} \arrayiii{x}{y}{z} + \mathbf{F}_\mathrm{drag}.
  \label{eq:3body-2}
\end{equation}
Next, we rewrite \eq{3body-2} in dimensionless form by normalizing lengths to Hill radii $R_h$ (see \eq{Rh}) and times to $\Omega_0^{-1}$.  The unit of velocity is then the Hill velocity, $v_h = R_h\Omega_0$.  It can be shown that in Hill units $GM = 3\Omega_0^2 R_h^3 = 3$.  Dropping the $z$-term as we will treat planar configurations only, \eq{3body-2}, in Hill units, reads
\begin{equation}
  \frac{d\mathbf{v'}}{dt'} = \arrayii{2 v_y' + 3x'}{-2v_x'} - \arrayii{3x'/r'^3}{3y'/r'^3} + \mathbf{F}'_\mathrm{drag},
  \label{eq:3body-3}
\end{equation}
where $\mathbf{F}'_\mathrm{drag}$ is related to $\mathbf{F}_\mathrm{drag}$ as $\mathbf{F}'_\mathrm{drag} = \mathbf{F}_\mathrm{drag}/R_h^2 \Omega_0$.

\subsection{The gas drag force}
The drag force, $\mathbf{F}_\mathrm{drag}$, can be expressed in terms of a stopping time $t_s$,
\begin{equation}
  \mathbf{F}_\mathrm{drag} = -\frac{\Delta \mathbf{v}}{t_s} = -\frac{(\mathbf{v}-\mathbf{v}_\mathrm{gas})}{t_s},
  \label{eq:Fdrag}
\end{equation}
where $\mathbf{v}_\mathrm{gas}$ is the velocity of the gas in the comoving frame and $\Delta \mathbf{v}$ the velocity difference between that of the particle and the gas, see \fg{sketch}. Due to pressure support, the gas rotates slower than Keplerian by a magnitude $v_\mathrm{hw} = \eta v_K$, where $v_K$ is the Keplerian velocity at disk radius $a$ and $\eta$ a dimensionless quantity that gives the fractional deviation from the Keplerian motion \citep{NakagawaEtal1986}:
\begin{equation}
  \eta = \frac{dP/da}{2a\Omega^2 \rho_g} \sim \left( \frac{c_g}{v_K} \right)^2,
  \label{eq:eta}
\end{equation}
with $c_g$ the sound speed.  In the comoving frame the headwind is directed towards negative $y$.  However, we should correct for the Keplerian shear; thus,
\begin{equation}
  \mathbf{v}_\mathrm{gas} = (-v_\mathrm{hw} - \frac{3}{2}\Omega x) \mathbf{e}_y,
  \label{eq:vgas}
\end{equation}
where $\mathbf{e}_y$ is the unit vector in the $y$ direction.

For the drag force we consider several regimes. The stopping time for solid spheres of internal density $\rho_s$ for particles of increasing size $s$ reads \citep{Weidenschilling1977}:
\begin{equation}
  t_s = \begin{cases}
    \displaystyle
    \frac{\rho_s s}{\rho_g c_g} & \textrm{(Epstein drag)} \\[5mm]
    \displaystyle
    \frac{4\rho_s s^2}{9\rho_g c_g \ell_\mathrm{mfp}} & \textrm{(Stokes drag)} \\[5mm]
    \displaystyle
    \frac{6\rho_s s}{\rho_g |\mathbf{v}-\mathbf{v}_\mathrm{gas}|} & \textrm{(Quadratic drag)} \\
  \end{cases}
  \label{eq:drag-regimes}
\end{equation}
where $\rho_g$ the density of the gas, and $\ell_\mathrm{mfp}$ the mean free path of the gas.  For small particles the Epstein regime holds.  The Stokes regime supersedes the Epstein regime for particle sizes $s>9\ell_\mathrm{mfp}/4$.  In both the Epstein and the Stokes regime the gas drag is linear with velocity and the stopping time reflects a particle property.  These are the regimes for which our study is applicable.  In the quadratic regime the stopping time becomes a function of the particle velocity since here $F_\mathrm{drag}\propto |\Delta\mathbf{v}|^2$.  In fact, there is a transition regime between the Stokes and quadratic drag regimes where stopping times are proportional to $|\Delta \mathbf{v}|^{0.4}$, which we have, for reasons of simplicity, ignored here (following \citealt{Rafikov2004}).  %This gives an error to the stopping time of at most a factor of 5.

As an (approximate) upper limit for $|\Delta \mathbf{v}|$ we can take the headwind velocity, $v_\mathrm{hw}$.  The transition between the Stokes and the quadratic drag regimes then occurs at a size of
\begin{align}
  \label{eq:smax}
  s \approx&\ s_\mathrm{max} = \frac{27 \ell_\mathrm{mfp} c_g}{2 v_\mathrm{hw}} \\ \nonumber
  =&\ 90\ \mathrm{m} \left( \frac{c_g}{10^5\ \mathrm{cm\ s^{-1}}} \right) \left( \frac{\rho_g}{10^{-10}\ \mathrm{g\ cm^{-3}}} \right)^{-1} \left( \frac{v_\mathrm{hw}}{30\ \mathrm{m\ s^{-1}}} \right)^{-1}, 
\end{align}
where we used $\ell_\mathrm{mfp} = 2\times10^{-9}/\rho_g$ (in cgs units; \citealt{NakagawaEtal1986}).  Since we consider a drag law that is linear in velocity, our results are only applicable for particle sizes less than $s_\mathrm{max}$.  In the inner disk (where the gas density is large) the results should be applicable to the sweepup of chondrule-like particles and m-size boulders.  In the outer disk, $\rho_g$ is much lower and the particles for which our results are applicable include (small) planetesimals.

%\begin{equation}
%|x| = \left\{ \begin{array}{rl}
% -x &\mbox{ if $x<0$} \\[2mm]
%  \frac{\displaystyle+2 \rho_s s}{\displaystyle \rho_g c_g} &\mbox{ otherwise}
%       \end{array} \right.
%\end{equation}

%The drag force then becomes
%In this study, we consider a linear dependence of the drag force on velocity, $F_\mathrm{drag} = -K_\mathrm{drag} (v-v_\mathrm{gas})$. 
%\begin{equation}
%  \mathbf{F}_\mathrm{drag} = K_\mathrm{drag} \arrayii{-v_x}{-v_y -\Delta v -3\Omega x/2}
%  \label{eq:Fdrag}
%\end{equation}
Expressed in dimensionless form the drag law reads
\begin{equation}
  \mathbf{F}_\mathrm{drag}' = \frac{\mathbf{F}_\mathrm{drag}}{v_h\Omega_0} = \frac{1}{\mathrm{St}} \arrayii{-v_x'}{-v_y' -\zeta_w -3x'/2}
\end{equation}
where we used \eq{Fdrag} for $\mathbf{F}_\mathrm{drag}$ and \eq{vgas} for $\mathbf{v}_\mathrm{gas}$, normalized velocities to $v_h$, and have introduced the Stokes number, $\mathrm{St}=t_s \Omega_0$,\footnote{Note that the Stokes number in this study simply indicates the dimensionless friction time; it is not necessarily the same as the Stokes number used in turbulent studies, $\mathrm{St}=t_s/t_L$ where $t_L$ is the turn-over timescale of the largest eddies.  For $t_L = \Omega^{-1}$ the definitions agree \citep{YoudinLithwick2007}.} and the dimensionless headwind velocity, $\zeta_w$:
\begin{equation}
  \zeta_w \equiv \frac{v_\mathrm{hw}}{\Omega R_h} \approx 12.5 \left( \frac{\rho_s}{\mathrm{g\ cm^{-3}}}\right)^{-1/3} \left( \frac{v_\mathrm{hw}}{30\ \mathrm{m\ s}^{-1}} \right) \left( \frac{R_p}{100\ \mathrm{km}} \right)^{-1} \left( \frac{a}{1\ \mathrm{AU}} \right)^{1/2},
  \label{eq:zeta-w}
\end{equation}
with $R_p$ the radius of the (proto)planet.  Note that due to the normalization to $v_h$, $\zeta_w$ is primarily an indicator of the size of the (proto)planet rather than of the strength of the headwind $v_\mathrm{hw}$ as the latter is approximately constant throughout the disk.

%An alternative way of writing the drag force on the test particle is in terms of the stopping time, $F_\mathrm{drag} = -(v-v_\mathrm{gas})/t_s$, with
%\begin{equation}
%  t_s = \frac{3}{4c_g \rho_g} \frac{m}{A},
%  \label{eq:tauf}
%\end{equation}
%\Refs\ where $A$ is the projected cross section of the particle.  For spheres of radius $s$ \eq{tauf} reduces to $t_s = \rho_s s/\rho_g c_g$. The dimensionless drag parameter then becomes $K'_\mathrm{drag} = (t_s \Omega) \equiv 1/\mathrm{St}$ with $\mathrm{St} = t_s \Omega$ the Stokes number.  Since we treat a linear drag law the formulation in terms of stopping times and Stokes number is especially advantageous since they express real particle properties that do not depend on velocity. 

\subsection{Dimensionless quantities}
\begin{figure}
  \centering
  \includegraphics[width=0.45\textwidth]{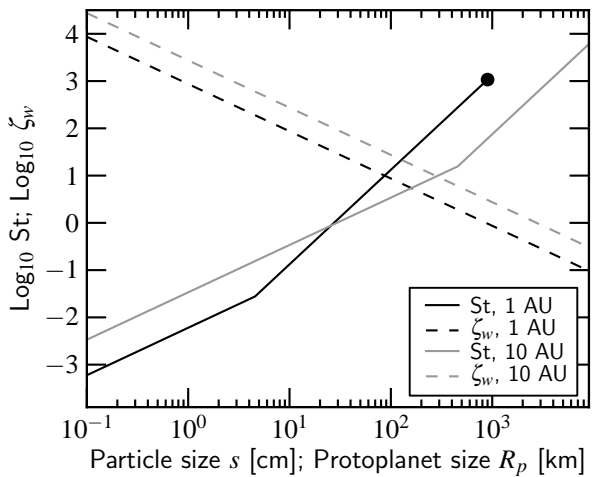}
  \caption{\label{fig:dim-quan}Relation between dimensionless and physical quantities.  The Stokes number ($\mathrm{St}$, \textit{solid} lines) and the dimensionless headwind velocity $\zeta_w$ (\textit{dashed} lines) are plotted on the $y$-axis as function of the radius $s$ of the particle and the radius $R_p$ of the protoplanet.  Note the different units of $s$ and $R_p$ on the $x$-axis.  The black dot denotes $s_\mathrm{max}$ (\eq{smax}).  Lines are shown for : \sumi\ $c_g=10^5\ \mathrm{cm\ s^{-1}}$ and $\rho_g=10^{-9}\ \mathrm{g\ cm^{-3}}$ at a disk radius of 1 AU (\textit{black} lines) and \sumii\ $c_g = 6\times10^4\ \mathrm{cm\ s^{-1}}$ and $\rho_g = 10^{-11}\ \mathrm{g\ cm^{-3}}$ at a position of 10 AU (\textit{gray} lines).  The internal density of solids is fixed at $\rho_s = 3\ \mathrm{g\ cm^{-3}}$, the nebula headwind is $v_\mathrm{hw}=30\ \mathrm{m\ s^{-1}}$, and the mass of the central object is solar.}
\end{figure}
\begin{table}
  \caption{\label{tab:dim-quan}Dimensional and dimensionless parameters.}
  \begin{tabular}{lll}
  \hline
  \hline
    Description               &  Dimensional  & Dimensionless \\
  \hline
    Hill radius               & $R_h$               & 1 \\
    Orbital frequency         & $\Omega$            & 1 \\
    Headwind velocity         & $v_\mathrm{hw}$     & $\zeta_w$ \\
    Drag constant             & $K_\mathrm{drag}$   & $K_\mathrm{drag}' = 1/\mathrm{St}$ \\  
    Radius (proto)planet      & $R_p$               & $\alpha_p$ \\
    Stopping time             & $t_s$            & $\mathrm{St}$ \\
    Collision rate            & $P_\mathrm{col}$    & $P$ \\
    Impact radii$^a$          & $b_\sigma$               & $b_\sigma$ \\
  \hline
  \end{tabular}
  \\[1mm]
  Note.|$^a$ For impact radii and velocities we intentionally use the same symbols, see also footnote \ref{foot:foot1}.
\end{table}
\Tb{dim-quan} compiles some key quantities in both dimensional and dimensionless form.  These include the dimensionless headwind velocity $\zeta_w$ (\eq{zeta-w}), the particle Stokes number $\mathrm{St}$, and the protoplanet radius 
\begin{equation}
  \alpha_p = \frac{R_p}{R_h} = 5.7\times10^{-3} \left( \frac{M_\star}{M_\odot} \right)^{1/3} \left( \frac{\rho_s}{\mathrm{3\ g\ cm^{-3}}} \right)^{-1/3} \left( \frac{a_0}{\mathrm{AU}} \right)^{-1},
  \label{eq:alpha-p}
\end{equation}
which mainly depends on semi-major axis $a_0$.  In \fg{dim-quan} we give the relation between the dimensionless $\zeta_w$ and $\mathrm{St}$ to the physical protoplanet size $R_p$ and particle size $s$ for two disk radii $a$.  We have adopted $v_\mathrm{hw} = 30\ \mathrm{m\ s^{-1}}$ and disk parameters that correspond (approximately) to a typical minimum mass solar nebula model \citep{Weidenschilling1977i,HayashiEtal1985}. 

The full set of equations of motions in dimensionless form, dropping the primes, reads
\begin{equation}
  \frac{d}{dt}\arrayii{v_x}{v_y} = \arrayii{2v_y +3x -3x/r^3}{-2v_x -3y/r^3} - \frac{1}{\mathrm{St}}\arrayii{v_x}{v_y+\zeta_w+3x/2}.
  \label{eq:eom}
\end{equation}
The drag-free equations of motions are retrieved when $\mathrm{St}\rightarrow \infty$, which signifies that particles are not coupled to the gas.  However, for Stokes number $\mathrm{St} \lesssim 1$ particles are coupled to the gas and the importance of the drag terms becomes relevant or even dominant. \citet{PetitHenon1986} and \citet{IdaNakazawa1989}, working in the gas-free regime, only had to care about the first terms on the RHS of \eq{eom} and the equations of motions did not include any parameter.  The addition of gas drag introduces two parameters: the velocity of the gas flow $\zeta_w$ and the coupling parameter $\mathrm{St}$.  Together with the size of the planet, $\alpha_p$, these fully specify the problem; \ie\ impact parameters $b_\sigma$ depend on these three dimensionless quantities only. Although not as clean as the drag-free equations, \eq{eom} still represent a significant reduction of the parameters involved (semi-major axis, particle size $s$, protoplanet size $R_p$, headwind velocity, gas density, \etc). In our parameter study we only have to care about these three parameters. 

We do not include the eccentricity in our prescription (and also not the inclination since the interaction is assumed to be 2D). Rather, the initial velocity of the approaching particle is given by the radial drift equations for individual particles, neglecting the 2-body interaction term.  These we will now review.

\section[]{The geometrical limit\footnote{\label{foot:foot1}In this and the next two sections lengths ($x,y,b_\sigma$, \etc) and velocities ($v$) are expressed in dimensionless (Hill) units, unless otherwise specified. } }
\label{sec:geo}
Ignoring the 2-body interaction force, we will analytically solve for the particle's trajectory in the comoving frame. As we will soon see, impact parameters along a particle trajectory are generally not conserved.  We provide a general relation between the impact parameter at the interaction point ($b_\sigma$) and its projected value on the $x$ and $y$ axes at any arbitrary point (\eq{xS}).  This relation will be used later in \se{orbitints} to obtain the impact radii $b_\sigma$.

\subsection{Steady-state velocities}
Without the two-body interaction term, the motion of the particle fulfills the well-known drift equations \citep{Weidenschilling1977,NakagawaEtal1986,BrauerEtal2007}:
\begin{subequations}
  \label{eq:vrphi}
  \begin{equation}
    v_r = - \frac{2v_\mathrm{hw} \mathrm{St}}{1+\mathrm{St}^2};
  \end{equation}
  \begin{equation}
    v_\phi = - \frac{v_\mathrm{hw}}{1+\mathrm{St}^2},
  \end{equation}
\end{subequations}
where $v_r$ is the radial velocity and $v_\phi$ the azimuthal velocity with respect to the \textit{local} Keplerian rotation.  Thus, in the context of a \textit{fixed}-rotating orbital frame we have $v_x = v_r$ and $v_y = v_\phi - \frac{3}{2}\Omega_0 x$, or in dimensionless units (divide by $v_h$)
\begin{subequations}
  \label{eq:vxy}
\begin{equation}
  v_x = - \frac{2\zeta_w \mathrm{St}}{1+\mathrm{St}^2}
  \label{eq:vx}
\end{equation}
\begin{equation}
  v_y = - \frac{\zeta_w}{1+\mathrm{St}^2} - \frac{3}{2}x
  \label{eq:vy}
\end{equation}
\end{subequations}
and it can be verified that with these expressions the RHS of \eq{eom} vanishes when the two body interaction terms ($-3x/r^3$ and $-3y/r^3$) are omitted.

\subsection{The parabola solution}
\begin{figure}
  \centering
  \includegraphics[width=\figw]{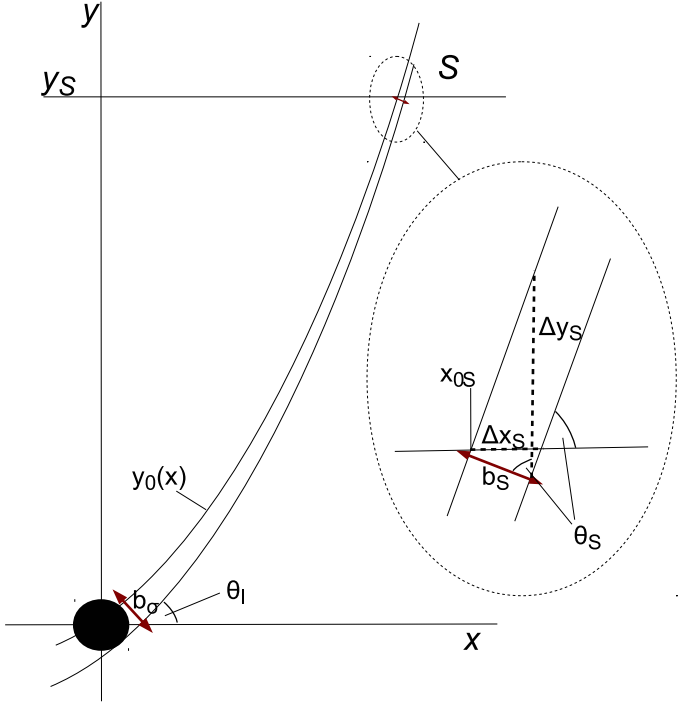}
  \caption{\label{fig:parabola}Without the two body force, particle trajectories as witnessed from the comoving frame obey parabolas. Two trajectories are shown: one that passes through the origin ($y_0(x)$) and one that just hits the target.  The corresponding impact parameter $b$ is denoted by arrows.  For curved trajectories $b$ is not conserved due to the changing slope of the curves, here indicated by the angle $\theta$.  At $S$ we have that $\Delta x_S = b_S/\sin \theta_S$ and $\Delta y_S = b_S/\cos \theta_S$.  $\Delta y_S$ is a conserved quantity. }
\end{figure}
Since
\begin{equation}
  \frac{dy}{dx} = \frac{v_y}{v_x} = \frac{1}{2\mathrm{St}} + \frac{3(1+\mathrm{St}^2)}{4\mathrm{St}\zeta_w}x
  \label{eq:dydx}
\end{equation}
we immediately recognize that the particle's trajectory in the rotating frame obeys a parabola, $y(x)=Ax^2 +Bx +C$ with $A =3(1+\mathrm{St}^2)/8\mathrm{St}\zeta_w$, $B=1/2\mathrm{St}$, and $C$ the integration constant, which is determined by the starting point $S$ of the particle.  We refer to the function that intersects the origin ($C=0$) as $y_0(x)$.  The starting point $(x_{0S}, y_S)$ is defined to lie on this curve, $y_S = y_0(x_{0S})$, see \fg{parabola}, where $y_0(x)$ is plotted by the upper parabola.  Another parabola with the same $A$ and $B$ (\ie\ for the same particle properties $\mathrm{St}$ and $\zeta_w$) but with non-zero $C$ is drawn in such a way that it just hits the `target' at the origin.  Its `launch point' at $y=y_S$ is shifted over a length $\Delta x_S$.  The vertical difference between the curves, $-C=\Delta y_S$, is preserved.

If the particle trajectories were straight, impact parameters would be the same everywhere in the $(x,y)$-plane.  However, due to the $x^2$-term this statement no longer holds for the general case of nonzero $A$.  See \fg{parabola}: the impact parameter near the origin or at the interaction region, $b_\sigma$, differs from that at $S$.   Impact parameters are no longer conserved due to the change in $dy/dx$.  The changing slope of the curve is indicated in \fg{parabola} by the angle $\theta$, where $\theta$ is related to \eq{dydx} as $\theta = \arctan (dy/dx)$. 

Using the properties of the parabola solution we can relate the quantities at $I$ to those at $S$.  At the interaction point $I$ the impact parameter is $b_\sigma$ (=$\alpha_p$ in the geometrical case) and the associated vertical width is $\Delta y_I = b_\sigma/\cos \theta_I$ with $\theta_I$ the angle the parabola makes at this point.  Similarly, $\Delta y_S = b_S/\cos \theta_S$ and due to the invariance of $\Delta y$ we therefore have that $b_\sigma = b_S \cos \theta_I/\cos \theta_S$.

The associated change in $x$ at the starting point is $\Delta x_S = \Delta y_S/\tan \theta_S$.  This can be expressed in terms of the impact parameter at the interaction point $b_\sigma$, \ie\
\begin{equation}
  \Delta x_S = \frac{\Delta y_S}{\tan \theta_S} = \frac{b_\sigma}{\cos \theta_I \tan \theta_S} = b_\sigma \frac{\sqrt{1+\left( dy/dx \right)^2_I }}{(dy/dx)^2_S},
  \label{eq:xS}
\end{equation}
where we used that $1/\cos \theta = 1/ \cos [\arctan (dy/dx)] = \sqrt{1+(dy/dx)^2}$. Of course, in the non-gravity limit we know already that the impact parameter is just the planet radius, $b_\sigma=\alpha_p$. However, \eq{xS} is essential to interpret the numerical result of \se{orbitints}. That is, in our numerical integration (that includes the 2-body force) we will scan the $x$-axis for trajectories that lead to a collision and obtain a length scale $\Delta x_S$ over which particles hit the target.  Using the above equation we can then relate the obtained range of projected impact parameters at $S$, $\Delta x_S$, to the impact parameter at the interaction point, $b_\sigma$.  

\subsection{Collision rates}
%\begin{figure}
%  \centering
%  \includegraphics[width=\figw]{./../Fig/CR/cr_geo.eps}
%  \caption{\label{fig:cr-geo}Collision rates (in Hill units) for several values of $\alpha_p$ and $\zeta_w$ corresponding to \eq{Pgeo}.  Multiply by $R_hv_h$ to convert to physical units. }
%\end{figure}
The collision rate $P$ in the 2D configuration is the product of the collision cross section, $2b_\sigma$, and the approach velocity, $v_a=\sqrt{v_x^2+v_y^2}$.  At $S$ we have that $b_S v_S = \Delta y_S v_x = \Delta x_S v_y$.  At $I$ the collision rate equals $b_\sigma v_a = \alpha_p v_a$.  Now, using \eq{xS} we have that $b_\sigma v_I = b_\sigma v_x/\cos \theta_I = v_x \tan \theta_S \Delta x_S = v_x \Delta y_S$  and we see that the rates at $I$ and $S$ are equal and \textit{independent of the choice for the starting point $y_S$}. This result just reflects mass conservation. Thus, we obtain the collision rate:
\begin{equation}
  P_\mathrm{geo} = 4\alpha_p \frac{\zeta_w \mathrm{St}}{(1+\mathrm{St}^2)} \sqrt{1 + \frac{(3\alpha_p (1+\mathrm{St}^2) +4\zeta_w)^2}{64\mathrm{St}^2 \zeta_w^2}},
  \label{eq:Pgeo}
\end{equation}
where $(dy/dx)$ has been evaluated at $x=\alpha_p$.  The complexity of \eq{Pgeo} may seem surprising for something as straightforward as a geometrical sweepup.  However, this is entirely due to the fact that \eq{Pgeo} covers several (velocity) regimes.  In \app{geolimit} we consider the asymptotic limits of \eq{Pgeo} and show that these correspond to the expected sweepup rates (cross section $\times$ approach velocity) and to the findings of \citet{KaryEtal1993}.

\section{Full 3-body integrations including gas drag and gravity}
\label{sec:orbitints}
\subsection{Description of the adopted algorithm}
We perform a parameter study of \eq{eom}, varying $\zeta_w$ and $\mathrm{St}$. For the dimensionless headwind velocity runs were performed at $\zeta_w = 0.01, 0.03, 0.1,\dots10^4$ and for the Stokes number values of $\mathrm{St}=10^{-4}, 3\times10^{-4}\dots10^4$ were sampled.  Thus, we obtain a grid of $17\times13=221$ different combination of $\zeta_w$ and $\mathrm{St}$.  Not every combination is equally likely. Indeed, the parameter space samples areas where our key approximations (linear drag law, constant gas density) lose validity,  but we intentionally sample a broad range of values to verify the validity of our analytical expressions (see \se{model}).

\begin{figure}
  \centering
  \includegraphics[width=\figw]{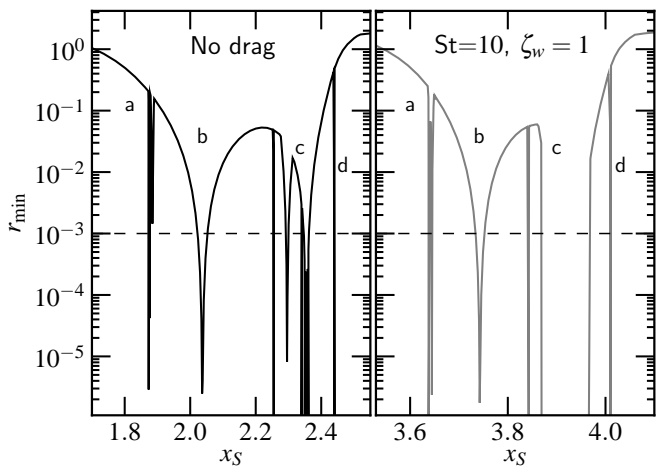}
  \caption{\label{fig:Rmin}The minimum distance in units of Hill radii to the origin (center of the protoplanet), $r_\mathrm{min}$, as function of the $x$-coordinate of the starting point, $x_S$ ($y_S$ is fixed at 40 Hill radii). Plotted is $r_\mathrm{min}$ for a gas-free system (\textit{left}) and a system that is characterized by the parameters $\mathrm{St}=10$ and $\zeta_w=1$ (\textit{right}).  Several bands are labeled.  The inclusion of gas drag shifts the bands to larger $x_S$ while merging several chaotic features within the chaotic c-band.}
\end{figure}
%Both directions are considered to find the minimum, \ie\ $y_S = \pm40$.  
%The motion of the particle is followed for $r<\mathrm{max}(50, r_S)$. 
For each combination of $\zeta_w$ and $\mathrm{St}$ we numerically determine the function $r_\mathrm{min}(x_S)$.  Particles are launched from a starting point $(x_S,y_S)$ where $y_S$ is fixed and $x_S$ is varied, see \fg{parabola}.  The initial velocities are given by \eq{vxy}. Depending on the sign of $v_y(x_S)$ the initial $y$-position ($+$ or $-$) is determined, such that the initial motion in $y$ is always directed towards the planet.  Here, we fix $|y_S|$ at 40 \citep{IdaNakazawa1989}.  Particles that leave the computational domain (when $|y|>40$ or $x<-40$) are no longer followed.  For a certain $x_S$ we numerically integrate \eq{eom} adopting an relative error of at most $10^{-8}$.  As our integrator we use a fifth-order Runge-Kutta scheme with timestep control \citep{Fehlberg1969,ShampineEtal1979}.   After the calculation has terminated we determine (and store) the minimum distance, $r_\mathrm{min}$.  In this way $r_\mathrm{min}(x_S)$ is obtained, see \fg{Rmin}.  Projected impact parameters $\Delta x_S$ are then obtained from the $r_\mathrm{min}(x_S)$ curve by summation over the intervals where $r_\mathrm{min} < \alpha_p$, \ie\
\begin{equation}
  \Delta x_S = \int dx_S H( \alpha_p - r_\mathrm{min}[x_S]),
  \label{eq:DxS}
\end{equation}
where $H(t)$ is the Heaviside step function,
\begin{equation}
  H(t) = \begin{cases}
    1 & t\ge0; \\
    0 & t<0.
  \end{cases}
\end{equation}
(Simply put: we only include the orbits that hit the target).

As $r_\mathrm{min}$ is occasionally found to vary steeply with $x_S$, fine sampling of the $x$-axis is required.  Therefore, we sample our parameters space ($x_S$) adaptively.  We start out with intervals of 1 Hill radii, \eg\, $\{x_S\} = 0, 1, 2, 3, \dots$.  In the next level the interval spacing is reduced by a factor 10, $\delta = 0.1$.  However, we only treat the points that fulfill the condition $r_\mathrm{min}(x_S) < F \delta$, where $F$ is empirically fixed at $10^3$ (see below).  For example, if $r_\mathrm{min}(0) = 300$ this point will be skipped in the next iteration of the algorithm.  If $x_S$ fulfills the condition, however, then both sides will be scanned; \eg\ if $r_\mathrm{min}(2) = 15 < 0.1F$ then, we will additionally perform calculations for $x_S=1.1,1.2,\dots 1.9$ and $x_S=2.1,2.2,\dots2.9$.  In this way we reduce the number of calculations but are still able to obtain a good assessment of $\Delta x_S$ for low $\alpha_p$.

Despite this optimization, we were forced to perform a relatively large number of integrations, \ie\ a large $F$.  The reason is the presence of very narrow, chaotic bands. In \fg{Rmin} band b near $x_S=2.0$ is a regular band since $r_\mathrm{min}$ varies smoothly with $x_S$.  A low $F$ value suffices to pick up this feature.  However, bands a and d are very narrow and show (if one would zoom in) additional substructure.  These chaotic bands are not resolved when choosing a low $F$.  In fact, there is no guarantee that our algorithm will pick up every band since they can be very narrow.  However, with $F=10^3$ we do obtain a good correspondence to previous works \citep{PetitHenon1986,IdaNakazawa1989}, also matching the substructure within the narrow bands shown in \fg{Rmin}.% so we have confidence in our method.  %Besides, the chaotic band contribute only little to the impact cross section.

Following the discussion in \se{geo} we emphasize again that the starting points ($x_S$ values) are not fundamental, but depend on the \textit{choice} of the starting point $y_S$.  Taking a different value of $y_S$, \eg\ $y_S=80$, will shift the features of \fg{Rmin}b towards higher $x_S$ values. In addition, the spacing (width of the features) will be different. The only conserved (physical) quantity is the mass flux, \ie\ the integral of $\int v_y(x_S)\mathrm{d}x_S$ over the width of the feature -- independent of the choice of $y_S$.

\begin{figure*}
  \centering
  \includegraphics[width=\textwidth]{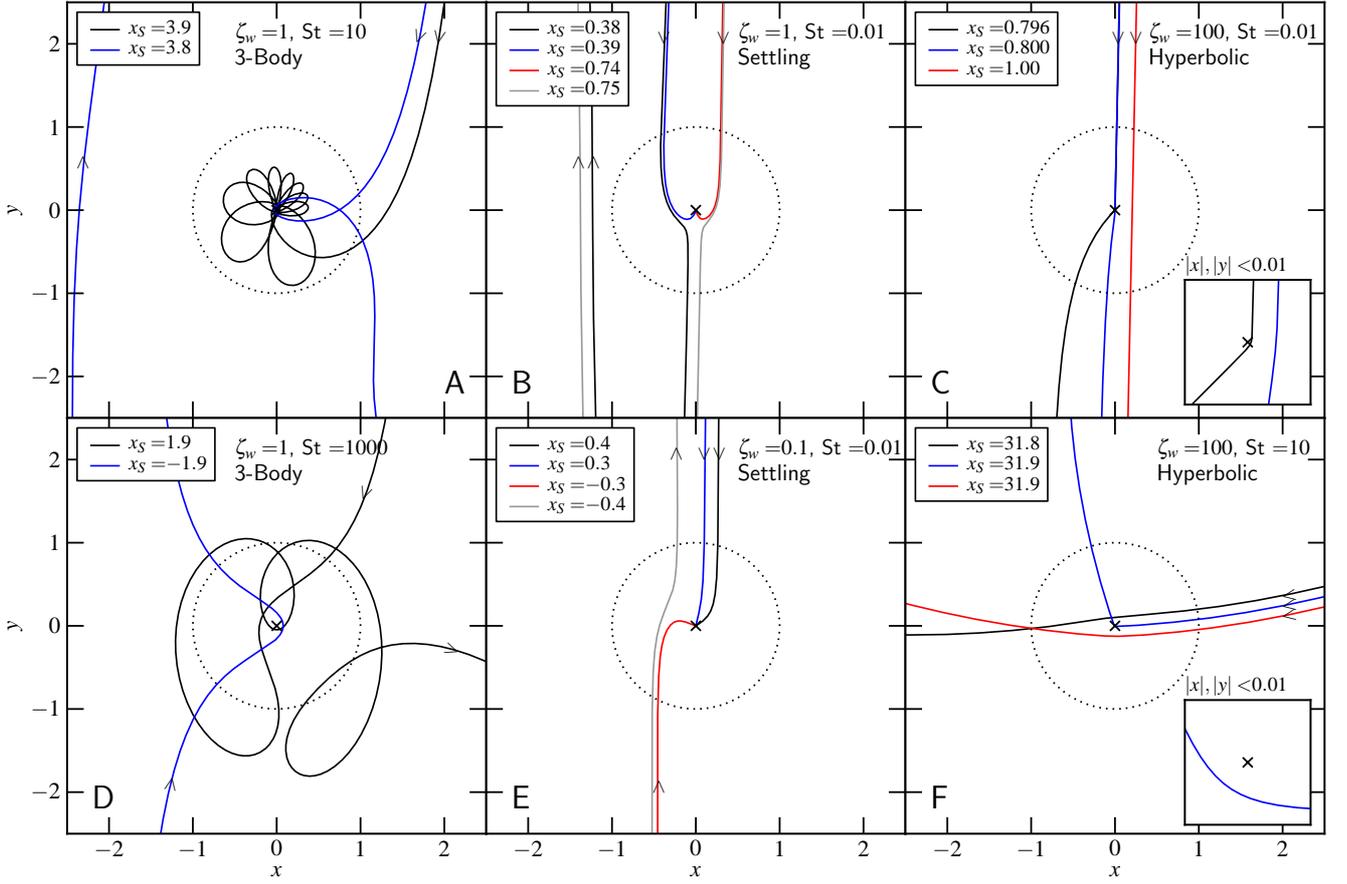}
  \caption{\label{fig:orbits}Examples of planet-particle interactions for different values of the dimensionless headwind velocity $\zeta_w$ and coupling parameter $\mathrm{St}$. For typical nebula parameters particles of $\mathrm{St}=10$ correspond to loosely coupled $m$-size particles, whereas $\mathrm{St}=0.01$ are more strongly coupled cm-size particles, see \fg{dim-quan}. Likewise, $\zeta_w=1$ corresponds to protoplanets of $R_p\sim10^3$ km in radius, while $\zeta_w=100$ corresponds to $R_p\sim10$ km planetesimals.  (A) Two particles of $\mathrm{St}=10$ experience a close encounter within the Hill sphere (dotted circle). The $x_S=3.9$ particle is captures and settles to the planet, whereas the other particle is ejected from the Hill sphere (The Keplerian shear eventually causes it to resurface at the other side of the Hill sphere). (B) Strong gas coupling, $\mathrm{St}=0.01$.  There is a competition between the gravitational pull of the planet and the drag force directed towards negative $y$.  (C) Close encounters at large $\zeta_w$ without settling (see inset). (D,E) Examples of particle trajectories originating from interior orbits.  (F) Radially approaching orbits. }
\end{figure*}
\subsection{Orbits including gas drag}
\Fg{orbits} provides several examples of particle trajectories that experience gas drag.  \Fg{orbits}a shows the trajectories for $\zeta_w=1$ and $\mathrm{St}=10$ with different starting points $x_S$ (\Fg{Rmin} contains the same parameters).  For a `standard' nebula setting, these parameters correspond to $\sim$m-size particles accreting onto a $\approx$$10^3$ km planet, see \fg{dim-quan}. Due to the large Stokes number, the influence of the gas is relatively weak and the orbits bear a close resemblance to the gas-free, three body regime.  The $x_S=3.8$ trajectory enters the Hill sphere, where it experiences a close encounter, at $r_\mathrm{min}=2.8\times10^{-2}$, before leaving the Hill sphere. It then re-emerges later at negative $x$ due to the combined effects of inwards radial drift and Keplerian shear.  However, the particle that started out at $x_S=3.9$ is captured within the Hill sphere and experiences strong orbital decay due to gas drag.

\Fg{orbits}b shows orbits for smaller particles of Stokes number $0.01$.  Four orbits are shown of which two lead to accretion.  Clearly, there is a contest between the gravitational pull of the planet and the aerodynamic pull of the gas flow.  Once close enough, gravity always wins. All orbits with $0.38\le x_S \le 0.74$ are accreted; there are no close encounters.  Accretion is independent of the physical proportion or internal density of the protoplanet; once a particle's angular velocity about the planet is damped by drag, it settles radially at its terminal velocity.  The only relevant physical quantity is the mass.  This mode of accretion reflects the capture mechanism of \fg{orbits}a.  We will refer to orbits like the $x_S=3.9$ curve in \fg{orbits}a as gas drag induced \textit{orbital decay}, whereas the accretion mode in \fg{orbits}b is referred to as \textit{settling} and draw the dividing line at $\mathrm{St}=1$.

On the other hand \fg{orbits}c, which features a larger dimensionless headwind (meaning: a smaller protoplanet) of $\zeta_w=100$, does not display the settling behavior.  Here, particles can only be accreted due to the finite size of the target.  The $x_S=0.796$ trajectory has a minimum distance of $r_\mathrm{min}=5.0\times10^{-4}$; the $x_\mathrm{S}=0.8$ trajectory $r_\mathrm{min}=4.5\times10^{-3}$.  Clearly, for a planet size $\alpha_p\ll1$ the impact parameter in \fg{orbits}c is much less than for the settling orbits of \fg{orbits}b.  Since the Stokes numbers are the same, the reason must be due to the larger headwind velocity $\zeta_w$.  This is understandable since particles of $\mathrm{St}\ll1$ approach at the headwind velocity ($v_a \approx \zeta_w$) and a large $v_a$ is not conducive for accretion.

In the lower panels of \fg{orbits} we vary either the Stokes number (particle size) or $\zeta_w$ (protoplanet size) with respect to the panel above.  For a Stokes number of $10^3$, see \fg{orbits}d, the effects of gas-drag are even less pronounced and it becomes more difficult to capture these (big) particles within the Hill sphere.  Moreover, if such a particle would be captured, it takes longer to finally accrete this particle due to orbital decay.  Another difference with \fg{orbits}a is that the $\mathrm{St}=10^3$ particles can now also enter the Hill sphere from \textit{interior} orbits (negative $x_S$).  In \fg{orbits}a the strong radial drift still prevents particles from entering the Hill sphere from the negative $y$-direction;  however, for $\mathrm{St}=10^3$ the radial drift is sufficiently reduced to render the situation more akin to the symmetric gas-free limit. 

The $\zeta_w=0.1$ orbits in \fg{orbits}e also feature accretion from particles approaching the planet from interior orbits, which the $\zeta_w=1.0$ orbits of \fg{orbits}b were not capable of.  The dimensionless headwind parameter of $\zeta_w = 0.1$ corresponds to a very big planet (in the canonical model) for which, as we will discuss below, the constant gas density background is unrealistic.  Alternatively, it can represent a smaller protoplanet in a nebula where the \textit{dimensional} headwind is, for some reason, strongly reduced.  In any case, we see that low $\xi_w$ tends to make the interactions more symmetric.  This can be seen from the $\eq{vy}$: low $\zeta_w$ or large $\mathrm{St}$ reduce the contribution from the non-symmetric headwind term, $\zeta_w/(1+\mathrm{St}^2)$.

\Fg{orbits}f shows, however, that for $\mathrm{St}=10$ and $\zeta_w=100$ the picture is anything but symmetric.  The particles approach the planet from a very radial direction ($x$-direction) -- at least, as seen from the perspective of the planet.  The point is here that both $\mathrm{St}$ and $\zeta_w$ are large.  Thus, both planet and particle move at a Keplerian velocity (in the azimuthal direction) but, due to the large $\zeta_w$, the particle still suffers a significant radial drift,  which outweighs the effects of the Keplerian shear.  As a result, the situation is similar to \fg{orbits}c: accretion does only proceed through close encounters.

\subsection{Collision rates}
We obtain the (dimensionless) collision rate from the encounters that hit the protoplanet, \ie\
\begin{equation}
  P(\alpha_p, \mathrm{St}, \zeta_w) = 2\int dx_S |v_y(x_S)| H(\alpha_p - r_\mathrm{min}[x_S]),
  \label{eq:Pcol}
\end{equation}
with $v_y(x)$ given by \eq{vy} and $H(t)$ the Heaviside step function. \Fg{cr} plots contours of $P(\zeta_w, \mathrm{St})$ for a planet size of $\alpha_p = 10^{-3}$, which corresponds to an heliocentric distance of $\approx$5 AU.  The reader must realize that $P$ is expressed in dimensionless units; the large rates that can be seen at large $\zeta_w$ (and $\mathrm{St}<1$) are less impressive upon multiplication by $R_h v_h \propto \zeta_w^{-2}$ (see \eq{zeta-w}).  In fact, these high $P$ values are consistent with the geometrical sweepup rates of \eq{Pgeo}.  However, the expression in terms of dimensionless units is useful since we can directly compare it to the gas-free limit for which $P_\mathrm{gf} \approx 11\alpha_p^{1/2} = 0.35$, see \se{gas-free}.  For large $\mathrm{St}$ and small $\zeta_w$, $P$ converges to $P_\mathrm{gf}$, the expected behavior.  However, for the remainder $P$ deviates significantly from $P_\mathrm{gf}$.  We sum up the main features:
\begin{enumerate}
  \item Particles of $\mathrm{St}\sim1$ accrete very well when the headwind velocity is low.  There is a distinct peak at $(\mathrm{St},\zeta_w) = (0,0)$; the accretion rate is here 20 times larger than \eq{dMdt}.  However, there is a very sharp transition between $1\lesssim \zeta_w \lesssim 10$.
  \item For large $\zeta_w$, $P=P_\mathrm{geo}$ is larger than $P_\mathrm{gf}$ although no gravitational focusing takes place.  The sweepup is so effective due to the strong headwind.
  \item For $\mathrm{St}>1$, the band $\mathrm{St} \sim \zeta_w$  features a maximum in $P$.
  \item For $\mathrm{St}\ll1$ and low $\zeta_w$ (large planets), collision rates are lower than $P_\mathrm{gf}$.  Tiny dust particles stay `glued' to the gas due to their strong coupling, preventing accretion.
\end{enumerate}
\begin{figure}
  \centering
  \includegraphics[width=\figw]{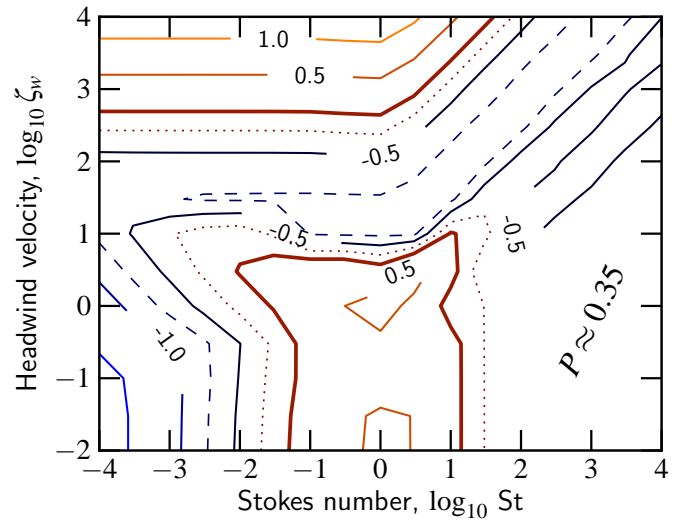}
  \caption{\label{fig:cr}Contour plot of collision rates obtained from the numerical integrations for $\alpha_p=10^{-3}$.  Contours of $\log_{10} P$ are shown as function of the Stokes number and the headwind velocity. Contour levels are indicated except for $\log_{10} P = 0$ (thick solid), $-0.25$ (dotted) and $-0.75$ (dashed). The accretion rate in the gas-free regime is $\log_{10} P_\mathrm{gf} \approx -0.46$.}
\end{figure}

\section{A simple model for gravo-gas interactions}
\label{sec:model}

\subsection{Model outline}
We present a simple model for the impact parameter $b_\sigma$.  The model is summarized in \fg{regimes}, where the three relevant regimes for the impact parameter are shown.  In the hyperbolic regime encounters follow the two-body approximation.  The usual gravitational focusing formula applies.  Keplerian shear is unimportant.  In the settling regime particles settle to the target and the impact parameter is independent of the planet size, $\alpha_p$.  For this reason, impact parameters can become rather large.  Finally, in the three-body regime, the encounter proceeds along the lines of the drag-free three body encounters at low energy.  However, the presence of the gas now causes some particles to be captured within the Hill sphere; these orbits decay and this enhances the accretion rate.
\begin{figure}
  \centering
  \includegraphics[width=\figw]{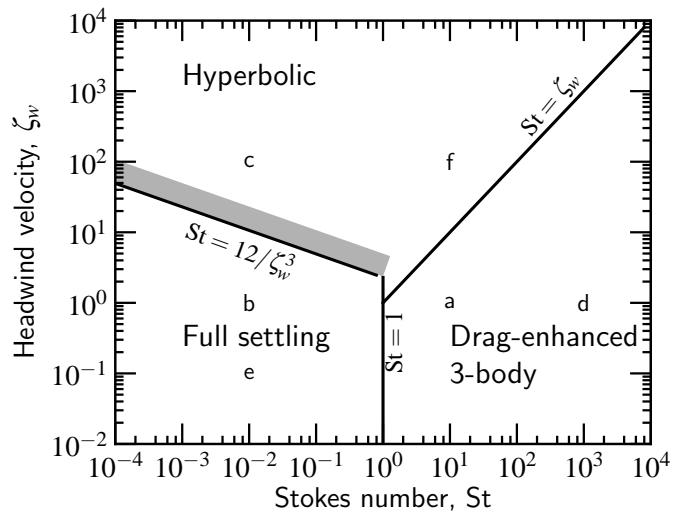}
  \caption{\label{fig:regimes}Illustration of the three accretion regimes.  In the hyperbolic regime interactions are 2-body encounters and the standard formula for gravitational focusing applies. In the settling regime, accretion proceeds through settling, which enhances the impact parameter $b_\sigma$.  Else, for $\mathrm{St}>\mathrm{max}(\zeta_w, 1)$ the solar gravity affects the encounter and $b_\sigma$ is increased with respect to the gas-free limit ($b_\mathrm{gf}$, \eq{bgf}) due to particle capture within the Hill sphere.  The gray band approximately indicates the zone where the settling solutions permeate into the hyperbolic regime. The letters a--f correspond to the parameters in the panels of \fg{orbits}.}
\end{figure}

\subsubsection{Importance of three body encounters.}
It is clear that for $\mathrm{St}\gg1$ the encounter cannot be described by a 2-body interaction, but should include the solar gravity.  But what is the transition between the 2-body and the 3-body regime in the presence of gas drag?  A passage through the Hill sphere typically takes a time of the order of the orbital period.  Thus, at first sight, we can draw the boundary at $\mathrm{St}=1$ since particles of lower stopping time will be strongly affected by the gas.  However, a large headwind velocity $\zeta_w$ has the same effect.  Particles that experience a drag force $\zeta_w/\mathrm{St}>1$ are blown out of the Hill sphere.  Thus, 3-body effects are reduced to the region of parameter space where $\mathrm{St}>1$ and $\mathrm{St}>\zeta_w$, see \fg{regimes}.

\subsubsection{Two body regime: settling- and hyperbolic interactions}
\label{sec:2breg}
We consider an interaction at impact parameter $b=b_\sigma$ at an approach velocity $v_a$.  The strength of the gravitational force is $f_g = 3/b^2$ and the interaction timescale, $t_a \simeq b/v_a$.  The latter quantity can be compared to the particle's response time $\mathrm{St}$. When $t_a<\mathrm{St}$ gas drag can be neglected during the encounter; the induced velocity change is $\Delta v = f_g t_a = 3b/v_a$.  However, if $t_a>\mathrm{St}$ the particle's velocity equilibrates towards $\Delta v = \Delta v_\mathrm{set} = f_g \mathrm{St} = 3\mathrm{St}/b^2$. 

For $\mathrm{St}<1$ the approach velocity can be approximately written as $v_a = 3b/2 +\zeta_w$.  For low $\zeta_w$ we therefore can expect settling since encounter timescales $t_a$ are long.  However, for large $\zeta_w$ settling will be prohibited: either the induced change $\Delta v$ is too little (at large $b$) or the interaction timescale too short for the particle to obtain its settling velocity (at low $b$).  

To see this quantitatively, the \textit{minimum} impact parameter for settling is $b^\ast = \zeta_w \mathrm{St}$ and the corresponding velocity change is $\Delta v^\ast = 3 /\zeta_w^2 \mathrm{St}$.  In order for the particle to settle to the central object, the direction of the particle has to change over a large angle, \ie\ $\Delta v \sim v_a$.  In fact, we obtain a better correspondence with our numerical result if we set the required velocity change to $v_a/4$.  Analytically, it can be shown that this is the required change for $\mathrm{St}\ll1$, see \app{path}. Thus, accretion through settling takes place when $v_a/4 \le \Delta v_\mathrm{set} \le \Delta v^\ast$ and disappears when $v_a/4 >\Delta v^\ast$.  At the boundary between the settling and hyperbolic regime it is allowed to take $v_a = \zeta_w$ (as can be verified \textit{a posteriori}).  We then have that for $\zeta_w^3 \mathrm{St} > 12$ settling is no longer possible, corresponding to a critical Stokes number
\begin{equation}
  \label{eq:Stcrit}
  \mathrm{St}^\ast = \frac{12}{\zeta_w^3},
\end{equation}
above which accretion through settling will no longer occur. 

\subsubsection{The settling regime}
Assuming the settling regime, particles at impact parameter $b$ experience a velocity impulse of $\Delta v_\mathrm{set} = 3\mathrm{St}/b^2$, which should equal $v_a/4$ for accretion. Since $v_a \approx 3b/2 + \zeta_w$ in this regime (note that we do not neglect the shear term since it becomes important at low $\zeta_w$) the condition $\Delta v_\mathrm{set} = v_a/4$ require us to solve the cubic equation
\begin{equation}
  b^3 + \frac{2\zeta_w}{3}b^2 - 8\mathrm{St} = 0.
  \label{eq:cubic}
\end{equation}
The (real, positive) solution to this equation is denoted $b_\mathrm{set}$.  %One can easily obtain more convenient approximations in the limits of small $\zeta_w, \mathrm{St}$ \etc. MORE TEXT HERE?

\subsubsection{Hyperbolic regime}
At large $\zeta_w$ the encounter is fast and the presence of gas drag can be ignored during the encounter.  For accretion we now require (by conservation of angular momentum) that $\Delta v = 3b/v_a \ge v_\mathrm{esc}$ which is much larger than in the settling regime.  For the impact radius we can just take the standard expression of the gravitationally-enhanced cross section,
\begin{equation}
  b_\mathrm{hyp} = \alpha_p \sqrt{ 1 + \left( \frac{v_\mathrm{esc}}{v_a} \right)^2} = \alpha_p \sqrt{1 + \frac{6}{\alpha_p v_a^2}},
  \label{eq:bhyp}
\end{equation}
where $v_\mathrm{esc} = \sqrt{6/\alpha_p}$ in Hill units.  For the approach velocity $v_a$ we now also include the horizontal velocity components (\ie\ $v_x$) since for $\mathrm{St}>1$ $v_x$ becomes dominant.  However, it is fine to neglect the shear term in \eq{vx} since $b_\mathrm{hyp}\ll1$.  Thus, for the approach velocity in the hyperbolic regime we can write
\begin{equation}
  v_\mathrm{a,hyp} = \zeta_w \frac{\sqrt{1 + 4\mathrm{St}^2}}{1+\mathrm{St}^2} %\approx \zeta_w\frac{1+2\mathrm{St}}{1+\mathrm{St}^2} \approx \zeta_w\frac{1}{1+\mathrm{St}/2}.
  \label{eq:bhyp-va}
\end{equation}

\subsubsection{The three body regime}
Without gas drag the effective impact parameter for collisions is \eq{bgf}, $b_\mathrm{gf} = 1.7 \alpha_p^{1/2}$, in dimensionless units (see \se{gas-free}).  Gas drag adds another component to the impact parameter on top of \eq{bgf}.  \Fg{Rmin} neatly illustrates this behavior.  The chaotic band c in the gas-drag simulation has collapsed.  Particles entering at the corresponding $x_S$-values are captured and decay to the central object on a timescale $\sim$$\mathrm{St}$.  If the gas inside the Hill sphere is removed within this timescale, these particles will become satellites; however, here we will simply assume that all captured particles contribute to the collision rate.

Because the accretion in the dissipative 3-body regime is determined by the behavior of the chaotic zones, it is difficult to provide an analytic model for the enhanced $b_\sigma$.  The chaotic zones are especially susceptible to collapse, because these particle trajectories are characterized by many revolutions, trough which a lot of energy can be dissipated.  In the gas-free situation one requires a positive energy $J$ to enter the Hill sphere,
\begin{equation}
  J = \frac{1}{2}v^2 -\frac{3}{r} -\frac{3}{2}x^2 +\frac{9}{2};
  \label{eq:J}
\end{equation}
and once $J$ becomes negative in the Hill sphere, \eg\ by inelastic collisions, the body becomes trapped \citep{Ohtsuki1993}.  Here, we face a similar situation where the gas drag is responsible for the energy removal.  Unfortunately, in our case an analysis in terms of the Jacobian is not so meaningful as the gas flow can also add energy; \ie\ $J$ is not conserved and bodies with $J<0$ can still be `blown out' of the Hill sphere.  However, the picture -- that gas drag can trap particles -- is still the key.

Empirically, we find that the impact radius is increased by a term $1/\mathrm{St}$, which corresponds to the dissipated energy over a revolution.  For these reasons, we add a term proportional to $1/\mathrm{St}$ to \eq{bgf},
\begin{equation}
  b_\mathrm{3b} = b_\mathrm{gf} + \frac{1.0}{\mathrm{St}} = 1.7\alpha_p^{1/2} + \frac{1.0}{\mathrm{St}},
  \label{eq:b3b}
\end{equation}
where the 1.0 constant is obtained empirically.  As we will see in the next section \eq{b3b} fits the general trend well, but it cannot reproduce the impact radius at every Stokes value.

\subsection{Comparison to numerical results and fine tuning of the recipe}
\begin{figure}
  \centering
  \includegraphics[width=\figw]{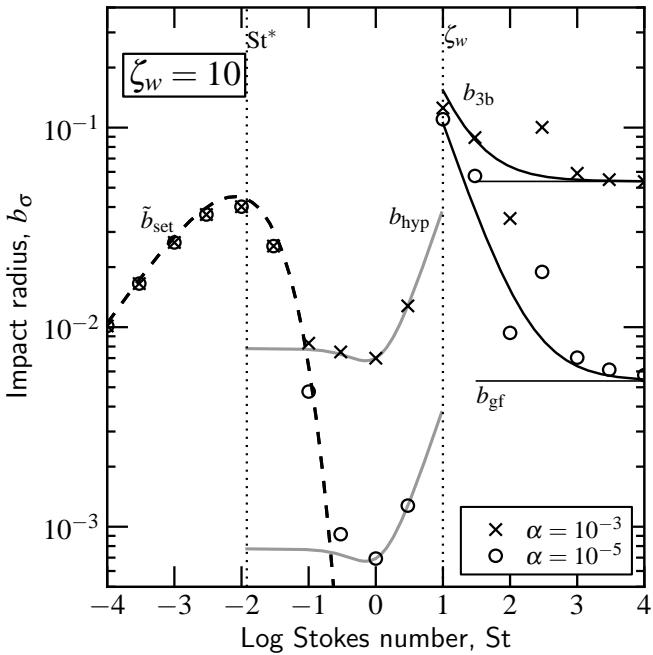}
  \caption{\label{fig:impactB}Impact radii from the numerical integrations (symbols) and analytic fits (curves) for a headwind velocity of $\zeta_w = 10$.  Vertical lines $\mathrm{St}^\ast$ and $\zeta_w$ distinguish the settling, hyperbolic, and three-body regimes.  Analytic fits from our recipe are shown by the dashed-black, solid-gray, and solid black curves, respectively, and denoted $\tilde{b}_\mathrm{set}, b_\mathrm{hyp}$ and $b_\mathrm{3b}$. The corresponding impact radii obtained from the numerical integrations are shown by \textit{crosses} ($\alpha_p=10^{-3}$) and \textit{circles} ($\alpha_p=10^{-5}$). }
\end{figure}
\Fg{impactB} compares the impact radii obtained from the numerical integrations (symbols) with the analytical prescriptions (curves) for a headwind velocity of $\zeta_w=10$ and for a planet size of $\alpha_p = 10^{-3}$ and $10^{-5}$.  We have used \eq{xS} to convert the projected impact parameters $\Delta x_S$ to true impact parameters at the interaction point ($b_\sigma$).  To do so we used the parabola solution, \eq{dydx}, to evaluate the gradients $(dy/dx)$ at the starting point ($x_S$) and at the interaction point $x_I$.  We determine the maximum value of $x_S$ that resulted in a collision with the planet at the specified $\alpha_p$ and took this value to compute $(dy/dx)_S$.  To compute $(dy/dx)_I$ we evaluated \eq{dydx} at the approach radius $b_\mathrm{app}$.  Here, for $b_\mathrm{app}$ we took the impact radius obtained from our analytical model described above, except for interactions in the 3-body regime where we always use $b_\mathrm{app}=2.5$.

At low $\mathrm{St}$ the interactions take place in the settling regime.  Impact radii are rather large, particularly near the  $\mathrm{St}^\ast = 12/\zeta_w^3$ transition line, and \textit{independent of $\alpha_p$} (the cross and circle symbols overlap), implying that the physical impact parameter is larger at larger disk radii.  For intermediate Stokes numbers the hyperbolic regime is valid and impact radii are much smaller. However, for $\mathrm{St}>\zeta_w$ impact radii once again increase.  The behavior is rather erratic, though, with peaks at $\mathrm{St}=10$ and 300 and a depression at $10^2$, valid for both $\alpha_p$. We found that this complex behavior can be attributed to the trajectories that originate from the third quadrant.  Initially, for low Stokes numbers, these are absent due to the strong radial drift.  However, at a critical Stokes number the contribution of particles approaching from interior orbits (negative $y_S$) becomes important.  We do not have a full understanding how these outliers can be modeled analytically.

The analytic fits to the various regimes are given by the dashed curve (for settling), solid gray curve (hyperbolic) and solid black curve (three-body).  From \fg{impactB} it is obvious that the transition between the settling and the hyperbolic regime is not so sharp.  Even particles that have $\mathrm{St}>\mathrm{St}^\ast$ display settling behavior.  For these reasons, we have extended the validity of the settling regime beyond $\mathrm{St}^\ast$ by adding an exponential term, \ie
\begin{equation}
  \tilde{b}_\mathrm{set} = b_\mathrm{set} \exp[-(\mathrm{St}/\mathrm{St}^\ast)^\gamma],
  \label{eq:bset-tilde}
\end{equation}
where $b_\mathrm{set}$ is the solution of \eq{cubic} and $\gamma$ a constant that we empirically fix at $\gamma=0.65$.  The impact parameter in the hyperbolic regime, $\mathrm{St}^\ast<\mathrm{St}<\zeta_w$, is then given by the maximum of $\tilde{b}_\mathrm{set}$ and $b_\mathrm{hyp}$. The little depression that can be seen at $\mathrm{St}=1$ is caused by the fact that the approach velocity \eq{bhyp-va} has a maximum here.  For larger $\mathrm{St}$ the approach velocity strongly decreases and the gravitational focusing strongly increases.  Nevertheless, impact parameters at $\mathrm{St}=\zeta_w$ are even larger than $b_\mathrm{hyp}$ and are better fitted by $b_\mathrm{3b}$.  Initially, gas drag very effectively captures bodies within the Hill sphere and impact radii are rather large.  However, the capture probability decreases as $1/\mathrm{St}$ and for large $\mathrm{St}$ we retrieve the gas-free limit, \eq{bgf}.
%The transition between the three regimes is rather sharp.  Our numerical results showed that settling also possible for $\mathrm{St}>12/\zeta_w^3$, albeit at a reduced rate.  We therefore extended the settling impact parameter by adding an exponential term to $b_\mathrm{set}$, \ie\ $\tilde{b}_\mathrm{set} = b_\mathrm{set} \exp[-(\mathrm{St}/\mathrm{St}^\ast)^{0.7}]$, where $\tilde{b}_\mathrm{set}$ is our new (more general) impact parameter. For the $b_\mathrm{hyp}, b_\mathrm{3b}$ transition (at $\mathrm{St}=\zeta_w$) we do not provide.

\subsection{Collision rates}
\begin{figure}
  \centering
  \includegraphics[width=\figw]{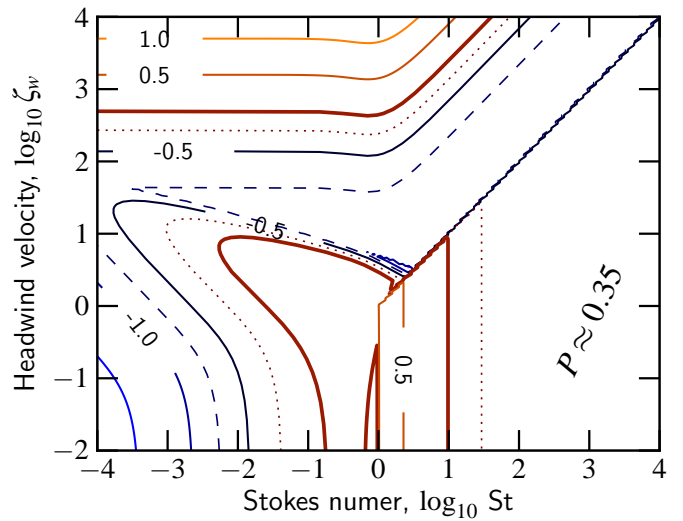}
  \caption{\label{fig:cr-th}Contours of $\log P$ according to the analytic prescription.  Curves are the same as in \fg{cr}.}
\end{figure}
In \fg{cr-th} we plot contours of the collision rate, that is, we plot $P=2b_\sigma v_a$ as function of $\mathrm{St}$ and $\zeta_w$ using the prescription outlined in \Tb{rec-sum}.  This figure should be compared with \fg{cr}.

The curves in \fg{cr-th} are much smoother due to the much finer grid that the analytic formulation permits.  The transition lines, $\mathrm{St}=\mathrm{St}^\ast$ and $\mathrm{St}=\zeta_w$ are clearly identified.  Our analytic formalism fails to reproduce the $\log P = -0.5$ band towards the upper-right of \fg{cr}.  However, the overall match is satisfactory; for 90\% of the 221 grid points the analytic and numerical results lie within 30\% of each other.

\subsection{Summary of impact parameter recipe}
\label{sec:sum-rec}
\begin{table*}
  \centering
  \caption{\label{tab:rec-sum}Summary of the analytic recipe to obtain the impact radii $b_\sigma$ and approach velocities $v_a$.}
  \begin{tabular}{p{5cm}lll}
    \hline
    \hline \\[-3mm]
    1. Calculate dimensionless parameters:&  $\zeta_w$ (headwind velocity)  & \eq{zeta-w}  \\
                                          &  $\alpha_p$ (planet size)       & \eq{zeta-w}  \\
                                          &  $\mathrm{St} = t_s \Omega$ (Stokes number)  & \eq{drag-regimes} \\
    \hline \\[-3mm]
    2. Calculate impact radii:  & $\tilde{b}_\mathrm{set}$        & \eq{cubic}, \eq{bset-tilde}    \\
                                & $b_\mathrm{hyp}$                & \eq{bhyp}     \\
                                & $b_\mathrm{3b}$                 & \eq{b3b}   \\
    \hline \\[-3mm]
    3. Determine Regime:                 & $\mathrm{St}<\min(1, 12/\zeta_w^3)$                  &                                 & $\mathrm{St}>\mathrm{max}(\zeta_w, 1)$ \\
                                          & Settling                                   & Hyperbolic                      & Three body \\
                                \cline{2-4} \\[-3mm]
    4. Results \\ 
    Impact radius (accretion), $b_\sigma$:  & $\mathrm{max}(\tilde{b}_\mathrm{set}, b_\mathrm{geo}$)  & $\mathrm{max}(\tilde{b}_\mathrm{set}, b_\mathrm{hyp})$    & $\mathrm{max}(b_\mathrm{3b}, b_\mathrm{geo})$ \\
    Approach velocity $v_a$:                & $3b_\sigma/2 + \zeta_w$                                    & \eq{bhyp-va}                    & $3.2$ \\
    Approach radius $b_\mathrm{app}$:       & $b_\sigma$                                      & $b_\sigma$                           & $2.5$ \\
    \hline
  \end{tabular}
  \flushleft
  Note.|Description of impact radii: $b_\mathrm{geo}$, geometrical impact radius ($=\alpha_p$); $b_\mathrm{set}$ impact radius in settling regime; $\tilde{b}_\mathrm{set}$, modified $b_\mathrm{set}$ (to cover the transition regime); $b_\mathrm{hyp}$ impact radius in the hyperbolic regime; $b_\mathrm{3b}$ drag-enhanced impact radius for the 3-body regimes; $b_\mathrm{app}$, approach distance.
\end{table*}
\Tb{rec-sum} provides an executive summary of how the collisional parameters can be obtained using the analytic prescription.  First, one converts the physical parameters (headwind velocity, disk radius, friction time, \etc) into the dimensionless quantities $\zeta_w, \alpha_p$ and $\mathrm{St}$.  The corresponding impact radii for the three regimes are calculated in the second step.  Then, in step 3, the appropriate collision regime is determined by comparing the Stokes number with $\mathrm{St}^\ast$ (\eq{Stcrit}) and $\zeta_w$ (\eq{zeta-w}), see also \fg{regimes}.  Dependent on the applicable regime, the final impact parameter is obtained by taking the maximum of two impact radii (step 4).  Other quantities ($v_a$ and $b_\mathrm{app}$) also depend on the collision regime.

Then, these results can be converted back to physical units by multiplication of $R_h$ and $v_h=R_h\Omega$ (\eq{Rh}) for, respectively, lengths and velocities.  The 2D-collision rate is then obtained from \eq{dMdt-2}.  The 3D-collision rate may be estimated by multiplication by a factor $\max(1, H_p/b_\sigma)$ (see \se{signif}), where $H_p$ is scaleheight of the particles.

\section{Significance to the growth of pre-planetary bodies}
\label{sec:signif}
In the previous sections we have outlined a general approach to analytically derive impact radii and collision rates in Hill coordinates.  But what does all of this imply for the growth of preplanetary bodies?  Perhaps the best way to illustrate this point is to calculate the accretion timescale
\begin{equation}
  T_\mathrm{ac}^\mathrm{2D} = \frac{M}{dM/dt} = \frac{4\pi \rho_s R_p^3/3}{P_\mathrm{col} \Sigma} = \frac{4\pi \rho_s R_p \alpha_p^2}{3 \Sigma P} \Omega^{-1},
  \label{eq:tac}
\end{equation}
where $P_\mathrm{col}$ is the dimensional accretion rate and $P=P_\mathrm{col}/R_h v_h$ the dimensionless, the quantity plotted in \fgs{cr}{cr-th}.  From \eq{tac} we see that the accretion timescale is inversely proportional to $P$ but also scales with $R_p$.  We further write \eq{tac} in terms of semi-major axis $a_0$ by substitution of \eq{alpha-p} for $\alpha_p$ and $\Omega(a)$ for a solar-mass star
\begin{equation}
  T_\mathrm{ac}^{2D} = \frac{6.7}{P} \left( \frac{\rho_s}{3\ \mathrm{g\ cm^{-3}}} \right)^{-1/3} \left( \frac{\Sigma}{1\ \mathrm{g\ cm^{-2}}} \right)^{-1} \left( \frac{R_p}{100\ \mathrm{km}} \right) \left( \frac{a}{\mathrm{AU}} \right)^{-1/2}\ \mathrm{yr}.
  \label{eq:tac2}
\end{equation}
The inverse dependence on disk radius may seem surprising but one has to realize that $P$ via $\alpha_p$ and $\zeta_w$ implicitly depends on $a$.  Nevertheless, \eq{tac2} shows that 2D accretion may be especially advantageous in the outer disks.

The 2D regime, however, may not be applicable to small particles since any breath of turbulence will stir them up.  The height of the particle layer may be obtained by equating particle diffusion and settling timescale; \ie
\begin{equation}
  \frac{H_p}{H_g} \approx \min\left(1, \sqrt{\frac{\alpha_t}{\mathrm{St}}} \right),
  \label{eq:Hp}
\end{equation}
\citep{DubrulleEtal1995,CarballidoEtal2006,YoudinLithwick2007} where $\alpha_t$ is the \citet{ShakuraSunyaev1973} viscosity parameter for turbulent diffusion.  In a 3D setting, the particle scaleheight can exceed the impact parameter $b_\sigma$; then, only a fraction, $b_\sigma/H_p$, of the particles take part in the interaction and the accretion timescale is correspondingly longer,
\begin{equation}
  T_\mathrm{ac}^\mathrm{3D} \approx T_\mathrm{ac}^\mathrm{2D} \times \max\left(1, \frac{H_p}{b_\sigma}\right)
  \label{eq:tac3}
\end{equation}
(with $b_\sigma$ in physical units). The 3D correction factor significantly increases collision timescales for small particles ($H_p$ is large) and the hyperbolic regime ($b_\sigma$ is small).

\begin{figure*}
  \centering
  \includegraphics[width=0.43\textwidth]{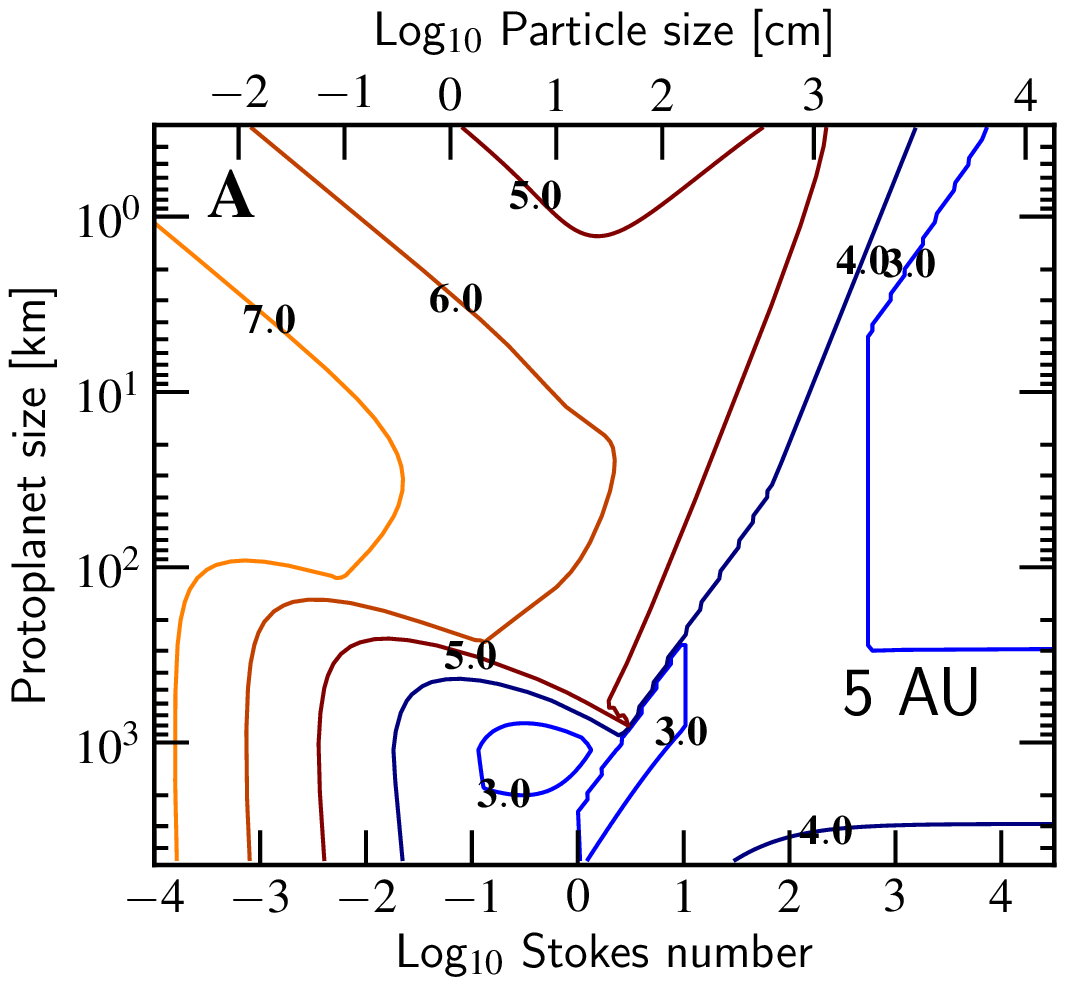}
  \includegraphics[width=0.43\textwidth]{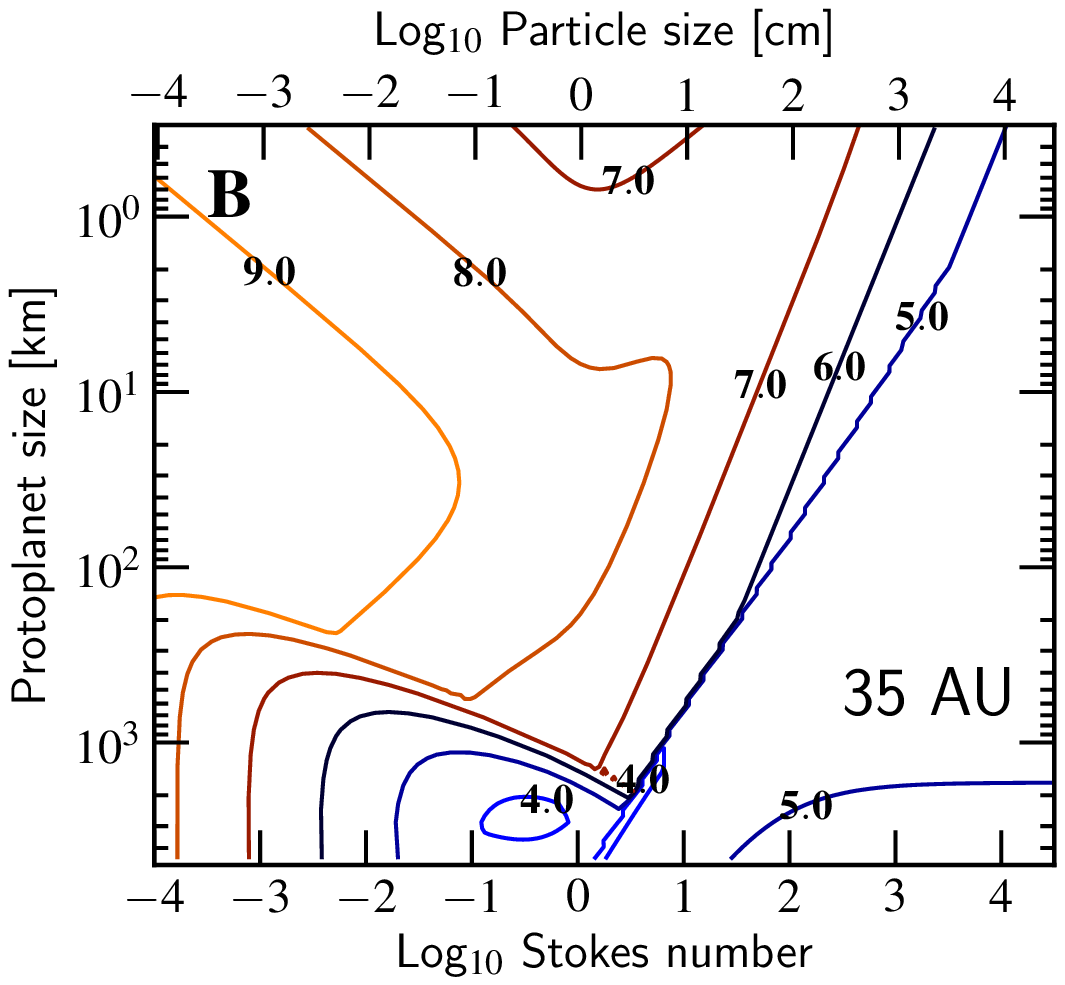}
  \caption{\label{fig:cr-ph}The 3D growth timescale $T_\mathrm{ac}^\mathrm{3D}$ as function of Stokes number (or particle size) and size of the protoplanet.   Contour lines of $\log_{10} T_\mathrm{ac}^\mathrm{3D}/\mathrm{yr}$ are shown.  All the solid density is assumed to be contained in particles of the indicated size.  (A) At 5 AU for a surface density of $\Sigma = 2\ \mathrm{g\ cm^{-2}}$. (B) At 35 AU for a surface density of $\Sigma=0.1\ \mathrm{g\ cm^{-2}}$.  In (A) the wider spacing between the tickmarks of the upper axis indicates particles enter the Stokes drag regime. }
\end{figure*}
In \fg{cr-ph}a we have plotted contours of the 3D growth timescales for a disk radius $a_0=5.2$ AU, $\rho_s=3\ \mathrm{g\ cm^{-3}}$ (making $\alpha_p = 10^{-3}$), $\Sigma = 2\ \mathrm{g\ cm^{-2}}$, $H_g = 0.25\ \mathrm{AU}$, $\Delta v_\mathrm{hw} = 30\ \mathrm{m\ s^{-1}}$, and $\alpha_t = 10^{-4}$.  The calculated accretion timescales assume that all the solid density is in particles of a single size.  Due the inclusion of the $H_p/b_\sigma$ factor the structure is quite different from that of \fg{cr-th}.  However, the contrast between the hyperbolic and settling regimes is still clearly visible and has in fact even increased due to the correction factor for the vertical structure.  Note that for the bigger bodies, which settle into a thin plane, \fg{cr-ph} still assumes that their eccentricities and inclinations are absent (low velocity regime).

\Fg{cr-ph} tells a few interesting points.  First, it can be clearly seen that growth of $\sim$km-size planetesimals by accretion of small particles ($\mathrm{St}<1$) takes a (perhaps prohibitively) long time.  Two mechanisms conspire.  First, the small particles couple effectively to the gas which dilutes their number densities near the midplane where the planetesimals are residing.  Of course, this statement depends on the strength of the turbulence that prevents the particles from settling effectively; timescales will be shorter for lower turbulent strength parameter, $\alpha_t$.  However, even in a completely laminar disk we may expect shear turbulence to develop \citep{Weidenschilling1980}, which strength may be equivalent to $\alpha_t$-values of $\sim$$10^{-6}$ \citep{JohansenEtal2006,CuzziWeidenschilling2006}.  Second, small particles, being strongly coupled, move with the gas, at a relative velocity of $\Delta v \approx v_\mathrm{hw} \gg v_\mathrm{esc}$, where $v_\mathrm{hw}$ is the velocity of the headwind and $v_\mathrm{esc}$ the escape velocity of the planetesimal.  Therefore, small particles lack gravitational focusing and it is hard to avoid the conclusion that sweepup of small particles by $\sim$km size planetesimals is slow.  In order to grow, planetesimals have to accrete among themselves.

However, the situation completely reverses when protoplanets sizes reach $\sim$10$^3$ km: for these bodies, accretion of cm to m-size particles becomes very rapid: in only $\sim$10$^3$ yr the protoplanet can double in size.  This is entirely due to the increased cross section in the settling regime. For `optimal' parameters ($\mathrm{St}\sim 1$, $\zeta_w\lesssim1$) the combined effect of gravitational focusing and gas damping results in impact parameters of $b_\sigma \sim 0.5R_h$ -- larger than what hitherto has been thought possible ($b_\mathrm{gf}$, see \eq{bgf}).  Accreting at impact parameters of the order of the Hill sphere is fast in any case but since $R_h$ increases with disk radii it is especially impressive for the outer disk, see \fg{cr-ph}b.  

For $\mathrm{St}\sim1$ particles, accretion is fast -- even though it is inefficient due to the strong radial drift.  We denote the probability that radially-inward drifting particles become accreted by the protoplanet $P_\mathrm{eff}$.  Since the drift flow is $2\pi a v_r \Sigma$ ($\approx$$0.15M_\oplus\ \mathrm{yr}^{-1}$ for $\mathrm{St}=1$ particles at 5 AU) $P_\mathrm{eff}$ is given as
\begin{equation}
  P_\mathrm{eff} = \frac{P_\mathrm{col}}{2\pi a v_r} \times \max\left(1, \frac{H_p}{b_\sigma}\right).
  \label{eq:Peff}
\end{equation}
(where we have again included the scaleheight correction factor). For a protoplanet of $R_p=10^3\ \mathrm{km}$ we find that $\mathrm{St}\sim1$ particles are accreted at an efficiency of only $P_\mathrm{eff}\approx 10^{-4}$.  The efficiency increases away from the $\mathrm{St}=1$ line ($v_r$ is lower) and towards larger protoplanet size and larger disk radii (larger $R_h$).  \citet{LevisonEtal2010}, using N-body techniques, also finds that accretion of small fragments was very inefficient (and concluded that it was therefore unlikely) except for a few specific particle sizes, that may have coincided with the peaks in \fg{cr-ph}.  To increase the accretion efficiency, smaller particles have to be accreted.  Dust fragmentation \citep{BirnstielEtal2009} or radial diffusion \citep{Ciesla2009} may be promising mechanisms to retain mm-size particles in the (outer) disk, where they are observed on $\sim$Myr timescales \citep[\eg][]{LommenEtal2009,RicciEtal2010}.

It is instructive to compare the accretion timescales of \fg{cr-ph}a to detailed hydrodynamical simulations involving $\mathrm{St}\sim1$ particles \citep{JohansenEtal2007,JohansenLacerda2010}.  In \citet{JohansenEtal2007} a dense particle layer of $\mathrm{St}\sim1$ boulders collapses into a Ceres-mass planet ($R\sim500$ km), that rapidly accretes the remaining boulders on timescales of perhaps 10 yr.  Although from \fg{cr-ph}a a Ceres-mass protoplanet in combinations with $\mathrm{St}\sim1$ particles form the optimal growth conditions, our accretion timescale of $10^3$ yr is still two orders of magnitude higher than what can be inferred from \citet{JohansenEtal2007}.  However, a direct comparison is perhaps not so meaningful since in the \citet{JohansenEtal2007} simulations the $\mathrm{St}\sim1$ particles are highly clumped and exert a strong feedback effect on the gas (\citealt{JohansenLacerda2010} discuss some alternate settings).  Feedback effects are not taken into account in this study. 

%There is one major caveat regarding the accretion of $\mathrm{St}\sim1$ particles.  The accretion may be fast, but due to the strong radial drift these particles experience, it is also very inefficient.  The accretion efficiency $P_\mathrm{eff}$, the probability that the particle gets accreted by the protoplanet during its inward drift, can be obtained by comparing $P_\mathrm{col}$ to the inward mass flow, $2\pi a v_r$; in 2D:
%But for the 3D case this is reduced by the same factor as in \eq{tac3}.  Of course, a large $P_\mathrm{col}$ reflects a large accretion efficiency $P$, but we still find that the impact probability for $\mathrm{St}=1$ particles at $R_p = 10^3$ km is $P_\mathrm{eff}<$0.1\%.  Thus, to grow this protoplanet by a factor of two in mass, one requires a mass in fragments equivalent to a solid body of $R_p=10^4$ km in size.  

%Once protoplanets reach $\sim$10$^3$ km proportions they have surely stirred up the planetesimal layer, rendering the zero eccentricity assumptions (on which \fg{cr-th} relies) implausible.  Indeed, in this oligarchic growth phase the growth is self-regulated, much slower than in the preceding runaway growth phase \Refs.  However, smaller bodies of $\mathrm{St}\lesssim1$ should not be affected by the protoplanet excitations since gas drag will quickly circularize their random motion.  Our results open up an avenue for extremely fast growth of protoplanets due to accretion of $\mathrm{St}\sim1$ particles by settling.

\section{Discussion}
\label{sec:discuss}
\subsection{Neglected effects}
We summarize the key assumptions that have been employed in this study:
\begin{itemize}
  \item a drag law linear in velocity;
  \item neglect of resonance trapping of particles;
  \item a smooth, laminar disk (only drift motions) without local pressure fluctuations;
  \item neglect of the gas flow around the protoplanet and of a possible atmosphere surrounding the proto(planet);
  \item a dynamically cold protoplanet on a circular, non-migrating, orbit.
\end{itemize}
The assumption of a linear drag law implies that this study -- and in particular the analytical prescriptions that have been derived -- apply for particles smaller than $\sim$$s_\mathrm{max}$ (see \eq{smax}) only.  But this still covers an appreciable size range, especially for the outer disk.

The adopted flow pattern in our study is unrealistic since it does not take account of the presence of the planet.  Of course, streamlines will have to bend around the object and this will affect the motion of the particle.  Tiny dust grains can only collide with a dust aggregate when the aggregate size is less than the mean free path of the gas molecules \citep{WurmEtal2001,SekiyaTakeda2005}.  However, this restriction probably applies only for small particles.  If we assume that the flowlines change over the lengthscale of the protoplanet, it follows that particles of $t_s < R_p/v_\mathrm{hw}$ are too tightly coupled to the gas to become accreted.  For example, for $R_p = 100$ km only $\mathrm{St}<10^{-4}$ particles are affected and our model overestimates the (already low) accretion rates.  More serious, perhaps, is our neglect of collective effects due to strong particle volume densities, as this will provide a feedback effect on the gas, affecting both the flow pattern as well as the drift rates.  This will probably be important for $\mathrm{St}\sim1$ particles and could significantly enhance the accretion rate (see our discussion at the end of \se{signif}).

For simplicity, our analysis only included drift motions.  We have, for example, neglected a systematic accretion (or decretion) flow of the gas.  In the turbulent $\alpha$-model this gas moves in at a velocity $\sim$$\alpha_t c_g H_g/a$.  Equating this expression with the radial drift velocity (\eq{vx}) we find that for particles $\mathrm{St}<\alpha_t$ the systematic accretion flow will dominate.  This will affect the expressions for the accretion rates.  Likewise, turbulent motions can become more effective to move particles around than drift motions, which affects the input parameter $v_a$ in our model described in \se{model}.  In the $\alpha_t$-model large eddies transport particles at velocities $\sim$$\alpha_t^{1/2}c_g$ \citep{CuzziWeidenschilling2006} and turbulent velocities will dominate the drift motions for $\alpha_t>\eta$ (see \eq{eta}).  The presented model may still be valid though, if the turbulent motions are included in the definition of the approach velocity, $v_a$.

Mean motion resonances may halt the particle long before it drifts to the Hill sphere \citep{WeidenschillingDavis1985}.  In such a situation the inward-directed drag force is balanced by the outward resonant perturbations. \citet{WeidenschillingDavis1985} showed that the strength of the perturbations is proportional to the planet's mass and to the resonance number $j$.  Thus, smaller particles (which experience a stronger drag force) move into a higher resonance.  However, this trend will not pursue indefinitely as at some maximum $j$ resonances will overlap and the effect is lost. \citet{Paardekooper2007} simulated particle accretion onto gas giants and showed that $\sim$m-size particles ($\mathrm{St}\sim1$) avoid resonance trapping for Jupiter-mass ($M_J$) planet.  For a planet of $0.1\ M_J$ the critical Stokes number has risen to $\mathrm{St}=10$ and for lower mass planets it will even be larger. Therefore, our results are not so much affected by resonance trapping when the protoplanet and core-formation stages are considered.

Our assumptions of a completely `inert' protoplanet is also peculiar.  In the oligarchic growth regime (where most of the mass resides in leftover planetesimals) dynamical friction will keep the eccentricities of the most massive bodies small \citep[\eg][]{KokuboIda2000}; however, if most of the mass is transferred to the protoplanets their motion will become eccentric (see \citealt{KaryLissauer1995} for accretion probabilities of protoplanets on eccentric orbits).  Likewise, type-I radial migration \citep{TanakaEtal2002} is not incorporated in our framework . These effects will again become important for already evolved protoplanets of masses $>0.1M_\oplus$.

The feedback effect of the protoplanet on the structure of the gas disk is also neglected.   The protoplanet's gravity influences the gas disk at larger distances, which could invalidate our approximation of a smooth (global) pressure.  Indeed, particles have a tendency to drift to high pressures regions and the $\mathrm{St}\sim1$ particles may be most affected by this process, piling up at a pressure bump instead of proceeding to the protoplanet.  \citet{Paardekooper2007} finds that this effect (together with the resonant trapping of bigger bodies) shuts off all accretion of particles sizes above $\sim$$10-100\ \mu$m!  However, this particle trapping is applicable for evolved planets only. \citet{MutoInutsuka2009} found that the criterion for particle trapping
\begin{equation}
  M_p \gtrsim \eta \frac{H_g}{a} M_\star \sim 10\ M_\oplus
  \label{eq:muto}
\end{equation}
for solar mass stars (\Eq{muto} is apparently independent of particle size or Stokes number).

Long before this size is reached, protoplanets bind the nebular gas and form atmospheres that will enhance the capture radius \citep{InabaIkoma2003,TanigawaOhtsuki2010}.  According to the results of \citet{InabaIkoma2003} this will perhaps become important when oligarchs reach $0.1 M_\oplus$.\footnote{We remark that the atmosphere calculations of \citet{InabaIkoma2003} do not take into account headwind flow, perhaps important for low $M_p$, which would destroy the spherical symmetry of the problem.}  Our expression for the impact radius and collision rates, therefore, are lower limits when protoplanets are surrounded by a thick atmosphere.

In summary, most of the mentioned effects become relevant for evolved (gas) planets only.  Most damaging to our analysis are the pressure fluctuations that could virtually shut off accretion, or make it very difficult to model it analytically.  However, this effect may only become effective for large planet masses (\eq{muto}).  For lower mass planets, the build-up of a dense atmospheres will enhance the accretion rates with respect to our prescription.  We believe that for dynamically-cold protoplanets below $0.1M_\oplus$ our prescription should be quantitatively correct if collective effects can be neglected.  In future work we intend to test the validity of our analytic expressions in more convoluted environments that incorporate some of the above processes.

\subsection{Summary}
We have developed a framework for the calculations of particle-protoplanet interactions in a gaseous environment.  This involves the integration of the equations of motions in the circularly restricted three body problem including drag forces.  Using the above mentioned simplifications -- most notably the assumption of a linear drag force, a smooth background density and headwind velocity, and a 2D setting -- we were able to reduce the general problem to a state that includes only two dimensionless parameters: the dimensionless headwind velocity $\zeta_w$ and the dimensionless stopping time (Stokes number, $\mathrm{St}$), see \se{sketch}.  A large parameter study of particle trajectories has been conducted, from which, as function of $\zeta_w$ and $\mathrm{St}$, the impact radii are derived.  We find that three accretion modes can be distinguished:
\begin{itemize}
  \item Settling encounters. Particles settle to the protoplanet and the impact radius is independent of the size of the latter;
  \item Hyperbolic encounters.  Accretion proceeds like the usual gravitational focusing with the approach velocity being influenced by gas drag.
  \item (Drag enhanced) three body encounters. Interactions take place on the scale of the Hill radius and gas drag causes a fraction of the particles to become captured, which settle to the protoplanet.
\end{itemize}
We have developed analytic recipes for all three encounters and found them to match very well to the results of our numerical study (except perhaps for the three-body regime).  The recipe is summarized in \Tb{rec-sum} and \se{sum-rec}.  In \se{signif} we have extended our approach to the usual 3D-setting and calculated the accretion times of (proto)planets by sweepup of particles.  We found that small particles are very unlikely candidates to grow to planetesimals of $10-10^2$ km, since their trajectories are tightly coupled to the gas.  However, if the protoplanet has reached a size of $\sim$10$^3$ km it can very quickly accrete $\mathrm{St}\sim1$ particles through the settling mechanism.  Since these fragments do not suffer from the protoplanet gravitational scattering (they quickly circularize), accretion under such conditions represents an avenue for quick growth, especially in the outer disk -- provided they are not lost by radial drift. 
%-- provided they have not been lost already by the radial drift. %with the provision that a large reservoir of fragments is available as the accretion efficiency is rather low.

\acknowledgements
The authors appreciate the helpful comments of the referee, Stuart Weidenschilling. C.W.O.\ acknowledges a grant from the Alexander von Humboldt Foundation and the hospitality of the Max-Planck-Institute for Astronomy for hosting him.  C.W.O.\ also appreciates stimulating discussions with Tilman Birnstiel, Kees Dullemond, Christoph Mordasini, and Marco Spaans.

\footnotesize
\bibliographystyle{aa}
\bibliography{ads,mybib,arXiv}
\normalsize

\appendix
\section{Asymptotic limits of \eq{Pgeo}}
\label{app:geolimit}
In this appendix we consider the asymptotic limits of \eq{Pgeo} and show the correspondence to the findings of \citet{KaryEtal1993}.

Three limits of \eq{Pgeo} can be identified
\begin{equation}
  P_\mathrm{geo} = \begin{cases}
    2\alpha_p \zeta_w              & \textrm{if } \mathrm{St} \ll 1; \\
    4\alpha_p \zeta_w/\mathrm{St}  & \textrm{if } \mathrm{St} \ll 1 \textrm{ and } \alpha_p \mathrm{St}/\zeta_w \ll 1; \\
    3\alpha_p^2                    & \textrm{if } \alpha_p \mathrm{St}/\zeta_w \gg 1. \\
  \end{cases}
  \label{eq:Pgeo-lim}
\end{equation}
These three regimes correspond to the cases where the square-root term of \eq{Pgeo} evaluates to $1/2\mathrm{St}$, 1, and $3\alpha_p\mathrm{St}^2/4\zeta_w$, respectively. In all limits we have assumed that $\alpha_p\ll\zeta_w$.%In the first two limits we have that the $\alpha_p$ term in the square-root of \eq{Pgeo} can be neglected.  In the first limit the square-root term becomes large (the constant 1 can be neglected); in the second limit the square root is close to unity.   In the third limit the $\alpha_p$-term dominates the square root.  

Rewritten in physical units \eq{Pgeo-lim} reads (\ie\ we multiply by $R_hv_h$) 
\begin{equation}
  P_\mathrm{col}^\mathrm{geo} = \begin{cases}
    2R_p v_\mathrm{hw}              & \textrm{for }\ \mathrm{St} \ll 1; \\
    4R_p v_\mathrm{hw}/\mathrm{St}  & \textrm{for }\ \mathrm{St} \ll 1 \textrm{ and } R_p \mathrm{St}/v_\mathrm{hw} \ll 1; \\
    3R_p^2\Omega                    & \textrm{for }\ R_p \mathrm{St}/v_\mathrm{hw} \gg 1. \\
  \end{cases}
  \label{eq:Pgeo-lim-phys}
\end{equation}
The interpretation of the first two limits is straightforward.  If $\mathrm{St}\ll1$ particles arrive with the headwind velocity, $v_\mathrm{hw}$.  In the 2D-setting the cross section is $2R_p$, so the collision rate is $2R_p v_\mathrm{hw}$.  Similarly, in the second limit the particles approach from the $x-$direction at a speed of $2v_\mathrm{hw}/\mathrm{St}$ (see \eq{vx}).  

In the third limit the particles approach once again from the $y$-direction.  However, for these very large particles,  the approach velocity is given by the Keplerian shear ($3\Omega x/2$) instead of the headwind, see \eq{vy}.  The approach velocity at the point of intersection, \ie\ at $x=R_p$, is then $v_a = 3R_p\Omega/2$.  Multiplied by $2R_p$ this reduces to the given expression.  The dependence on $R_p$ may seem counter intuitive but is natural in situations that involve shear.

The study of \citet{KaryEtal1993} concerned massive particles (\ie\ the third limit of \eq{Pgeo-lim-phys}).  \citet{KaryEtal1993} gave an analytic expression for the \textit{impact probability} or efficiency $P_\mathrm{eff}$ of a particles while crossing the semi-major axis of the protoplanet due to radial drift:\footnote{Note that we give the dimensional form.  Equation (7) of \citet{KaryEtal1993} is expressed in dimensionless units (but not in Hill units). }
\begin{equation}
  P_\mathrm{eff} = \frac{3R_p^2 \Omega^2}{16\pi K a v_\mathrm{hw}},
  \label{eq:Peff-Kary}
\end{equation}
where $K$ is the drag constant for a drag law that quadratically depends on velocity.

In our case $P_\mathrm{eff}$ can be found by taking the ratio of the collision rate $P_\mathrm{col}$ to the mass inflow rate $|2\pi a v_x|$ of the particles.  For the third limit of \eq{Pgeo-lim-phys} we obtain
\begin{equation}
  P_\mathrm{eff}^\mathrm{geo} = \frac{P_\mathrm{col}^\mathrm{geo}}{2\pi a v_x} = \frac{3 \mathrm{St} \Omega R_p^2 }{4\pi a v_\mathrm{hw}},
  \label{eq:Peff-geo}
\end{equation}
where we used that $v_x = 2 v_\mathrm{hw}/\mathrm{St}$ in this regime. \Eq{Peff-geo} is different from \citet{KaryEtal1993}'s result for three reasons:
\begin{enumerate}
  \item \citet{KaryEtal1993} considers a drag force quadratic in velocity where particles move at a drift velocity $v_x = -2K v_\mathrm{hw}^2/\Omega$ ($K$ has units $\mathrm{cm}^{-1}$).  However, we can mimic the linear drift law by substitution of $K=\Omega/v_\mathrm{hw} \mathrm{St}$ (\cf\ \eq{vx}) into \eq{Peff-Kary},.
  \item In our approach we have not accounted for the variation of the approach velocity over the impact range.  More correctly, the mean approach velocity is $\overline{v_a}=3\Omega R_p/4$.% (Note that for the three-body regime we do apply an averaged approach velocity of $3.2v_h$.)
  \item \citet{KaryEtal1993} do not take account of the planetesimals coming from the negative $y$-direction (the third quadrant), as they (correctly) argue that any such body should already have impacted during its approach from the first quadrant.  This is due to the highly symmetrical setting in this limit, which our naive reasoning above does not account for.  Equation \eq{Peff-geo} is therefore too large by a factor of two.
\end{enumerate}
With these corrections \eqs{Peff-Kary}{Peff-geo} agree. %In our analytic expressions (\se{model}) we have not accounted for the last two correction factors as we found them to have a marginal influence only.

\section{The settling path}
\label{app:path}
In the limit of $\mathrm{St} \ll 1$ we can use the approximation that the particle is always in the settling regime, \ie\ its velocity is given by $\mathbf{v} = \mathbf{F}_g t_s$ (in Hill units):
\begin{subequations}
  \label{eq:vset}
\begin{equation}
  v_x = -\frac{3x}{r^3} \mathrm{St},
\end{equation}
\begin{equation}
  v_y = -\frac{3y}{r^3} \mathrm{St} -\zeta_w.
\end{equation}
\end{subequations}
Thus, the particle path obeys the differential equation
\begin{equation}
  \frac{dy}{dx} = \frac{v_y}{v_x} = \frac{y}{x} + \frac{\zeta_w}{3\mathrm{St}} \frac{(x^2+y^2)^{3/2}}{x},
  \label{eq:dydx-set}
\end{equation}
which is slightly simplified if expressed in angular coordinates by the substitution $y=x \tan \theta$.  Then, $dy/dx = \tan \theta + x(1 + \tan^2 \theta) d\theta/dx$ and \eq{dydx-set} becomes in terms of $x$ and $\theta$
\begin{equation}
  x (1+\tan^2 \theta) \frac{d\theta}{dx} = \frac{\zeta_w}{3\mathrm{St}} (1+\tan^2 \theta)^{3/2},
\end{equation}
which is equivalent to
\begin{equation}
  \cos \theta \frac{d\theta}{dx} = \frac{\zeta_w}{3\mathrm{St}} x.
\end{equation}
Straightforward integration gives the solution
\begin{equation}
  \sin \theta = \frac{\zeta_w}{6\mathrm{St}} x^2 +C,
\end{equation}
with $C$ the integration constant, which we obtain by the requirement that at $\theta = \pi/2$ the particle starts out at $x=x_0$.  Thus, the full solution to the orbit under these conditions is
\begin{equation}
  1-\sin \theta = \frac{\zeta_w}{6\mathrm{St}}(x_0^2 -x^2).  
  \label{eq:setsol}
\end{equation}
During the encounter $\theta$ decrease from $\pi/2$ to $-\pi/2$ and $1-\sin \theta$ increases from 0 to 2.  The particle's $x$ coordinate then decreases by an amount that depends on the numerical value of $\zeta_w/6\mathrm{St}$ and $x_0$.  The particle settles to the origin if $x=0$ can be reached.  The largest value of $x_0$ for which this is possible is $b = x_0 = \sqrt{12 \mathrm{St}/\zeta_w}$.  This is the impact parameter $b_\sigma$.

For example, for $\mathrm{St}=10^{-2}$ and $\zeta_w=1$, we obtain $b_\sigma=0.35$, a value that is reasonably close to the numerically derived $x_S = 0.38$ (see \fg{orbits}).  For lower Stokes number the agreement becomes better.

In \se{2breg} we have used the expression $\Delta v = 3\mathrm{St}/b^2$ for the impulse change in the settling regime.  For $b=\sqrt{12 \mathrm{St}/\zeta_w}$ this corresponds to a velocity change of $\Delta v = \zeta_w/4$.  We further argued that for settling encounter the approach velocity $v_a$ should be changed by an amount $\Delta v \sim v_a$.  For small Stokes values $v_a=\zeta_w$.  Therefore, for $\mathrm{St}\ll1$ the criterion for accretion by settling becomes $v_a \ge \zeta_w/4$.   In \se{2breg} we have applied this criterion generally.

\end{document}